\documentclass[twocolumn]{aastex63}

\usepackage{multirow, graphicx, epsfig, color}
\DeclareGraphicsExtensions{.eps}

\shorttitle{Variability Signatures of Relativistic Shocks}
\shortauthors{B\"ottcher \& Baring}

\def\teq#1{$\, #1\,$}                         

\font\fiverm=cmr5

          \font\sixrm=cmr6       

\def\dover#1#2{\hbox{${{\displaystyle#1 \vphantom{(} }\over{
   \displaystyle #2 \vphantom{(} }}$}}
\def\ThetaBfone{\Theta_{\hbox{\sixrm Bf1}}}
\def\gammax{\gamma_{\rm max}}
\def\pmax{p_{\rm max}}
\def\ecyc{\omega_{\hbox{\fiverm B}}}
\def\wp{\omega_{\hbox{\sixrm p}}}

\begin{document}

\title{\rm \uppercase{Multiwavelength Variability Signatures of Relativistic Shocks in Blazar Jets}}

\author{Markus B\"{o}ttcher}
\affil{Centre for Space Research, North-West University, Potchefstroom, 2531, South Africa}
\email{Markus.Bottcher@nwu.ac.za}
\author{Matthew G. Baring}
\affil{Department of Physics and Astronomy - MS 108, Rice University,
6100 Main Street, Houston, Texas 77251-1892, USA}
\email{baring@rice.edu}

\begin{abstract}
Mildly-relativistic shocks that are embedded in colliding magnetohydrodynamic flows 
are prime sites for relativistic particle acceleration and 
production of strongly variable, polarized multi-wavelength emission from relativistic 
jet sources such as blazars and gamma-ray bursts. The principal energization
mechanisms at these shocks are diffusive shock acceleration and shock drift acceleration. 
In recent work, we had self-consistently coupled shock acceleration and radiation transfer simulations 
in blazar jets in a basic one-zone scenario. These one-zone models revealed that the observed 
spectral energy distributions (SEDs) of blazars strongly constrain the nature of the shock layer hydromagnetic 
turbulence. In this paper, we expand our previous work by including full time dependence and 
treating two zones, one being the site of acceleration, and the second being a larger emission
zone. This construction is applied to multiwavelength flares of the flat spectrum radio quasar 
3C~279, fitting snap-shot SEDs and generating light curves that are consistent with
observed variability timescales. We also present a generic study for the  
typical flaring behavior of the BL Lac object Mrk 501. The model predicts correlated
variability across all wavebands, but cross-band time lags depending on the type of
blazar (FSRQ vs. BL Lac), as well as distinctive spectral hysteresis patterns 
in all wavelength bands, from {\sl mm} radio waves to gamma-rays.  These 
evolutionary signatures serve to provide diagnostics on the competition 
between acceleration and radiative cooling.
\end{abstract}


\section{Introduction}
 \label{sec:intro}

Extragalactic jets of active galactic nuclei (AGN) and gamma-ray 
bursts (GRBs) are some of the most powerful emitters of radiation 
in the Universe, identifying them as sites of efficient particle 
acceleration. Relativistic, oblique, magnetohydrodynamic (MHD) 
shocks internal to these jets have long 
been considered as one of the leading contenders for the sites of 
relativistic particle acceleration that seeds the observed rapidly 
variable, often highly-polarized multi-wavelength (MW) emission. The 
dominant particle acceleration mechanisms at such shocks are
diffusive shock acceleration (DSA) and shock drift acceleration (SDA), 
inextricably linked and often collectively referred to as first-order Fermi 
acceleration.  

In DSA, particle energization results 
from repeated shock crossings of particles. For this process to be effective,
the particles's gyrational motion along large-scale ordered magnetic fields
has to be reversed by some process. In the case of DSA, this reversal of 
particle momenta $\mathbf{p}$ along magnetic field lines is facilitated by diffusive 
pitch-angle scatterings (PAS) in strong, chaotic MHD turbulence.
Note that many such pitch angle scatterings arise per gyroperiod, 
and their accumulation generates diffusive mean free paths that are usually fairly close 
to (but exceed) a charge's gyroradius -- see \cite{SB12} for details.  It is 
also possible that larger angle scatterings can contribute, an element that 
forms the focus of the blazar/shock acceleration study in \cite{Stecker07ApJ}.

In SDA, the gradient in the electric field across the shock discontinuity 
does work on charges and accelerates them promptly and in episodes of 
gyrational reflection off the shock layer, interspersed with upstream diffusive 
excursions by the particles in which they are forced to return to the shock 
by the dominant convective flow
\citep{DV86ApJ,SB12}. In contrast to DSA, for shock drift energization
to be at its most effective, the MHD turbulence level has to be relatively low,
so that reflections in the shock layer are not disrupted, and the net 
diffusive mean free path far exceeds the gyroradius.
Accordingly, DSA and SDA complement each other in terms of their acceleration capability, 
respectively dominating when the field turbulence is strong (DSA) near the shock discontinuity,
or the field is substantially more laminar on much larger spatial scales (SDA). This picture of 
a concentration of MHD turbulence nearer the shock is represented in Fig.~2 of BBS17.

We remark that the concept of SDA is dependent on the frame of reference. 
In the de Hoffman Teller (HT) shock rest frame, where the flow velocity ${\bf u}$ and 
magnetic field ${\bf B}$ vectors are parallel to each other both upstream and downstream 
(${\bf u} \times {\bf B} = {\bf 0}$), there is no large-scale electric field and, thus, no SDA. 
If one views the shock in the so-called normal incidence frame (NIF), which is 
obtained via a particular Lorentz boost ${\bf v}$ in the plane of the shock, 
the upstream flow velocity ${\bf u}$ is normal to the shock plane and 
a large-scale ${\bf v} \times {\bf B}$ electric field exists in the NIF, thereby facilitating 
an identifiable SDA.
For an extensive discourse on the relationship between plasma turbulence, charge transport and 
acceleration by the DSA and SDA processes, the reader may consult \cite{SB12} and \cite{BBS17}.

Theoretical studies of particle acceleration at relativistic shocks 
\citep[e.g.,][]{KH89,Ellison90,ED04,SB12} have shown that the shock 
acceleration process can result in a wide variety of spectral indices 
for the particle distribution, up to a limiting slope of \teq{n(p) \propto
p^{-s}} with \teq{s =1}. In particular, \cite{SB12} 
highlight the fact that flat \teq{s \sim 1} power laws  
develop when turbulence is low and SDA dominates the acceleration process, 
as charges are effectively trapped for long periods in or upstream of the shock layer.
These circumstances contrast the steeper distributions with \teq{s \sim 2.5} that 
emerge from particle-in-cell (PIC) kinetic plasma simulations where the 
Weibel instability enhances the turbulence that drives the acceleration process 
\citep[e.g.,][]{Sironi09ApJ,Sironi13ApJ},
but diminishes the trapping of charges near the shock layer.
Similar indices \teq{s \sim 2.2} are observed for electrons in PIC simulations 
of the current-driven Bell instability at mildly-relativistic shocks \citep{Crumley19MNRAS}. 
The reader can consult \cite{Marcowith16RPPh} 
for a comprehensive review of the microphysics of shock acceleration.
Note that magnetic reconnection models can also develop 
distributions with \teq{s\sim 1-1.5} \citep[e.g.][]{Cerutti12},
though PIC simulations of charge transport between X-point locales 
for energization and moving magnetic islands indicate a steepening of the 
acceleration distribution index to \teq{s\sim 1.5 - 4}, depending on the 
plasma magnetization \citep{Sironi14ApJ}. 

Such studies of particle acceleration usually 
do not consider the resulting radiative signatures in a self-consistent manner, 
and clearly observational constraints on \teq{s} and other byproducts of 
acceleration theory would be extremely insightful.
On the other hand, models focusing on the time-dependent, multi-zone radiative 
transfer problem for internal-shock models of blazars 
\citep[e.g.,][]{MG85,Spada01,Sokolov04,Mimica04,SM05,Graff08,BD10,JB11,Chen11,Chen12}
typically approximate the results of shock acceleration by assuming an ad-hoc 
injection of purely non-thermal relativistic particles, usually with a broken and/or
truncated power-law distribution in energy.  Therefore, blending these two 
aspects of the jet dissipation problem to enable deeper insights is strongly motivated.

To this end, in recent work \citep[][hereafter BBS17]{BBS17}, we coupled the 
Monte Carlo (MC) simulations of shock acceleration from \cite{SB12} 
with the steady-state radiative transfer routines of \cite{Boettcher13}. This 
provided, for the first time, a consistent description of 
the separate but intertwined mechanisms of
DSA and SDA and their radiative signatures in 
mildly relativistic, oblique shocks in blazar jets. An integral element of such 
an approach is that it includes complete distributions of leptonic populations of non-thermal
plus thermal particles, thereby enabling observational constraints on the values 
of important jet plasma quantities such as the lepton number density \teq{n_e}, 
and consequently the electron plasma frequency \teq{\wp = [4\pi n_e e^2/m_e]^{1/2}}
and the magnetization \teq{\Sigma = \sigma/\gamma = B^2/[\, 4\pi n_e m_ec^2\, ]}.
Fits to the spectral energy distributions 
(SEDs) of three blazars indicated the need for a strongly energy-dependent PAS diffusive 
mean-free path $\lambda_{\rm pas} \propto p^{\alpha}$, with $\alpha \sim 2$ -- 3, depending 
on the type of blazar. This may be considered as evidence of hydromagnetic turbulence levels 
gradually decreasing with increasing distance from the shock (BBS17),
and the dominance of SDA for electrons at energies exceeding $\sim 30$~MeV.

In this work, we present an extension of the shock acceleration + radiation-transfer model of BBS17, including
full time variability. We make predictions for time-dependent snap-shot SEDs, and produce MW light
curves, which can be further analyzed to predict multi-wavelength spectral hysteresis patterns and inter-band
time lags. A brief summary of our model and its application to a gamma-ray flare of the Flat
Spectrum Radio Quasar (FSRQ) 3C~279 in late 2013 and early 2014, which exhibited a negligible 
change of the Compton dominance compared to the quiescent state, was outlined in \cite{BB19}. 
 Here we present the full model description, and apply our model to another flare of 3C~279 which 
is part of the same active phase of this blazar in the 2013 -- 2014 epoch, but exhibits a greatly increased 
Compton dominance, as is more typical of the multi-wavelength flaring behaviour of FSRQs. 
Additionally, we detail two case studies for the BL Lac 
object Mrk 501. These are not applied to any specific flaring episodes, such as the 
stunning VHE variability on timescales of a few minutes reported in \cite{Albert07}, but 
rather to a generic description of the collection of flares garnered in MW campaigns for Mrk 501 
over the last two decades.

In Section~\ref{sec:scheme}, we describe our model setup and the numerical scheme we developed for 
simulating combined time-dependent shock acceleration and radiation transfer 
in internal shocks in blazars. Results of the application 
of our numerical scheme to two well-known $\gamma$-ray blazars are presented in Section~\ref{sec:results}. 
Specifically, we model two contrasting multi-wavelength flares of the FSRQ 3C~279 (Section \ref{sec:3C279}), 
one with an extreme increase of the Compton dominance, yielding 
good MW spectral fits and distinctive temporal characteristics illustrated using hardness-intensity diagrams 
and discrete correlation functions.  
An obvious strength of our 3C 279 modeling is that it simultaneously describes {\it both}
the multi-wavelength spectroscopy and the variability patterns.
We further present in Section~\ref{sec:Mrk501} template models of typical 
multi-wavelength flaring behaviour of the high-frequency-peaked BL Lac object (HBL) Mrk 501, 
with predictions of expected spectral variability behavior. Specifically, we model two test
cases: one in which the flaring is caused only by a change of the total power of particles accelerated
due to shock acceleration, and one representing the characteristic extreme synchrotron peak shift to
higher frequencies often observed in Mrk~501 during flaring states, requiring a significant change of 
the mean free path for pitch-angle scattering of shock-accelerated particles.  We summarize and discuss 
our results in Section~\ref{sec:discuss}. 
Throughout the manuscript, unprimed symbols denote quantities in the emission-region (jet) rest
frame, while a superscript `$\ast$' refers to quantities in the AGN rest frame and a superscript
`obs' signifies the observer's frame.

\section{Model setup and numerical scheme}
 \label{sec:scheme}

The plasma in relativistic jets of AGN is known to propagate at bulk speeds $\beta^{\ast}_{\Gamma}c$
corresponding to bulk Lorentz factors $\Gamma^{\ast} \sim 5$ -- 40 \citep[e.g.,][]{DG95,Jorstad05}. In the 
case of blazars, these jets are oriented at a small angle $\theta_{\rm obs}^{\ast} \lesssim 1/\Gamma^{\ast}$
to our line of sight, resulting in strong Doppler boosting of the emission by a Doppler factor 
of $\delta = 1/\sqrt{1 - \beta^{\ast}_{\Gamma} \, \cos\theta^{\ast}_{\rm obs}} \sim \Gamma^{\ast}$ 
in observed frequency and a factor of $\delta^4$ in observed bolometric flux, compared to quantities measured 
in the co-moving frame of the jet plasma. 

Our underlying assumption throughout this work is that mildly relativistic shocks 
with jet frame Lorentz factors \teq{\Gamma_{\rm s} \sim 1-3} propagate through the 
jets of blazars at all times, leading to time-variable diffusive shock acceleration in small 
acceleration zones proximate to shock fronts.  These modest \teq{\Gamma_{\rm s}} shocks 
naturally arise when two ultra-relativistic MHD flows collide. A quiescent state is established 
through a balance between continuous and steady particle energization in the acceleration 
zone, and radiative cooling and escape of particles in a 
larger radiation zone of length $\ell_{\rm rad}$
(measured in the co-moving frame of the jet material), 
which is identified with the high-energy emission 
region (see BBS17 and Figure~2 therein for details).  Enhanced emission and variability arises from 
the passage of a mildly relativistic shock through the density and magnetic field structures in the 
high-energy emission region, on an observed time scale $\Delta t_{\rm obs} = (\ell_{\rm rad} / v_{\rm s}) \, 
(1 + z) / \delta$.  Here $v_{\rm s}$ is the shock velocity in the co-moving frame of the jet material 
and $z$ is the cosmological redshift of the source.  Turbulence on larger scales that may 
well seed such shock structures is routinely generated in both hydrodynamic 
and MHD simulations \citep[e.g.,][]{Meliani08AA,Porth13MNRAS,BarniolDuran17MNRAS}.

In the conventional shock acceleration scenario, the first-order Fermi acceleration process 
that includes episodes of shock drift 
energization is facilitated by stochastic pitch angle diffusion of charges spiraling along magnetic field lines. 
A useful parameterization of the mean-free path for pitch angle scattering
is \teq{\lambda_{\rm pas}= \eta (p) \, r_{\rm g}}, i.e. via a momentum-dependent multiple 
\teq{\eta (p)} of the particle's gyro radius, \teq{r_{\rm g} = p c / (q B)}, where $p$ is the particle's momentum. 
A broadly applicable choice for the scaling is a power-law in the particle's momentum, \teq{\eta (p) = \eta_1 \, (p/mc)^{\alpha - 1}}, 
where \teq{\eta_1} describes the mean free path in the non-relativistic 
limit, \teq{\gamma \to 1}.  Motivations for this form from hybrid plasma simulations, quasi-linear MHD turbulence 
theory and {\it in-situ} spacecraft observations in the heliosphere are discussed in \cite{SB12,BBS17}.

For the acceleration/injection pipeline of our MW modeling here, an array of representative
thermal plus non-thermal particle distributions resulting from DSA+SDA for various values 
of the shock speed $v_{\rm s}$, magnetic-field obliquity \teq{\ThetaBfone}, and PAS mean-free-path parameters
$\eta_1$ and $\alpha$ have been generated using the Monte Carlo code of \cite{SB12}, with some examples 
being displayed in Fig.~1 of BBS17.  This ensemble of MC simulations illustrates that shock acceleration 
leads to a 
non-thermal broken power-law tail of relativistic particles 
which have been accelerated out of the remaining thermal pool.  
As a consequence of the \teq{\eta (p) \propto p^{\alpha -1}} form, 
the particle distribution is somewhat steep 
(\teq{dn/dp \sim p^{-2.2}}) at low momenta when DSA dominates, and much flatter
(\teq{dn/dp \sim p^{-1}}) for much higher momenta when SDA is the more effective energization process.
We note that these distributions are somewhat anisotropic in the shock rest frame 
(which moves  at a mildly-relativistic speed relative to the jet frame) due to the strong convective action in 
relativistic shocks -- e.g., see Figs.~4 and~5 in \cite{SB12}.

A high-energy cut-off 
at Lorentz factor \teq{\gammax \approx \pmax /m_ec}
of the non-thermal particle spectra results from the balance of the acceleration 
time scale $t_{\rm acc} (\gammax) = \eta (\gammax) \, t_{\rm g} (\gammax)$ 
with the radiative energy loss time scale. If synchrotron losses dominate, $\gammax \propto B^{-1/2}$. 
This will lead to a synchrotron peak energy $E_{\rm syn} \sim 240 \, \delta /[\, \eta (\gammax)\, ]$~MeV. 
Notably, this synchrotron peak energy is independent of the magnetic field $B$, as $E_{\rm syn} \propto 
B \, \gammax^2$. 
Blazars typically show synchrotron peaks in the IR to soft X-rays. In order to
reproduce these, the pitch angle scattering mean-free-path parameter \teq{\eta (\gammax)} has to 
assume values of $\sim 10^4$ -- $10^8$, first noted by \cite{IT96}.
However, \cite{SB12} have shown that $\eta_1$ must be significantly 
smaller than this \teq{\eta (\gammax)} value in order to obtain efficient injection of particles 
out of the thermal pool into 
the non-thermal acceleration process. From these arguments one can infer that \teq{\eta(p)} must be strongly 
dependent on momentum $p$ (BBS17).  

\begin{figure}
\vspace{0.5cm}
\centerline{\includegraphics[width=8.5cm]{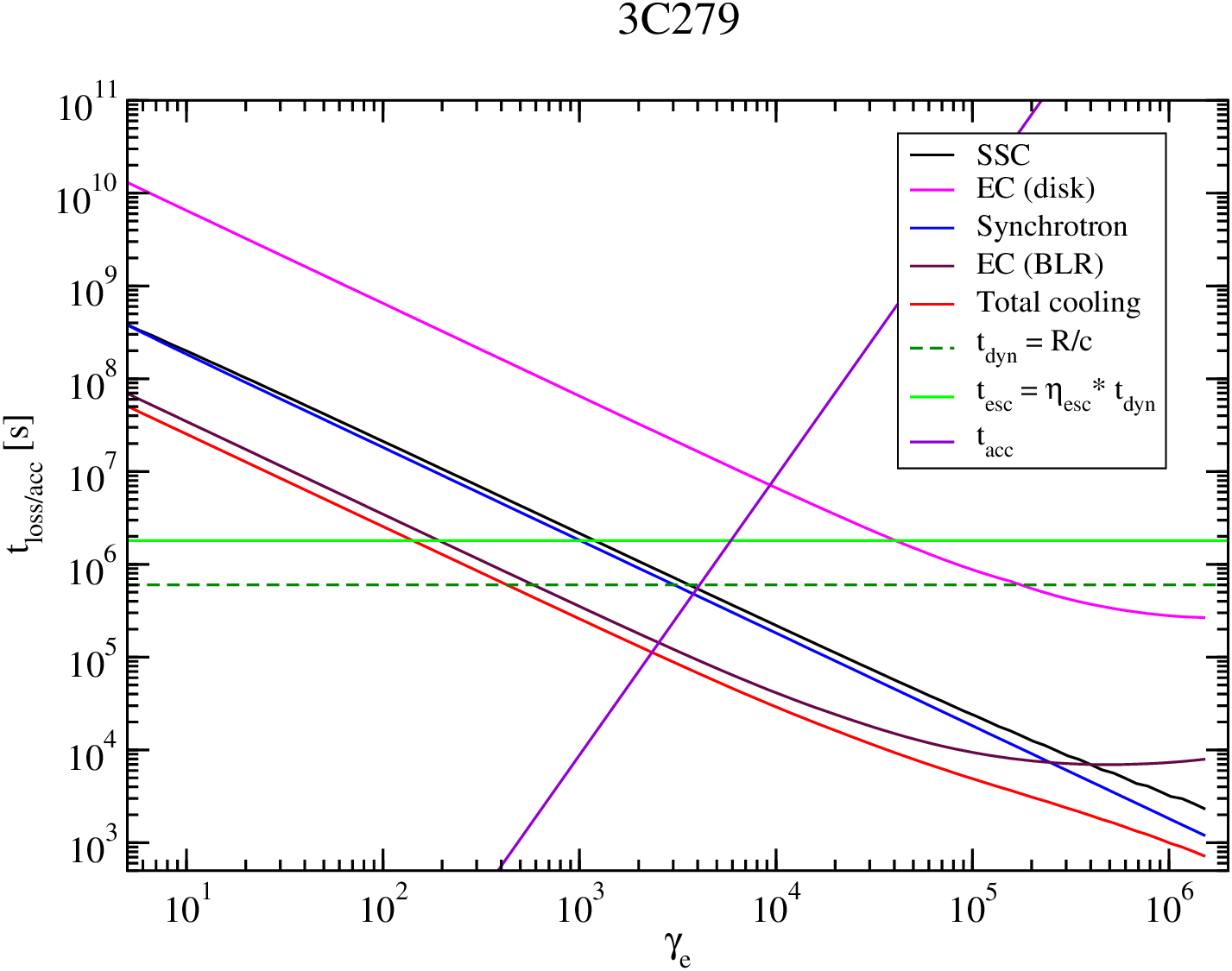}}
\vspace*{-5pt}
\caption{The relevant timescales as functions of electron Lorentz factor in the 
simulated quiescent-state equilibrium configuration for 3C~279, i.e. the green spectrum 
in Figure~\ref{3C279_SEDs_FlareC} (see Section \ref{sec:3C279} for details).  
The diagonal curves/lines represent the acceleration time (purple), 
radiative cooling (black, pink, blue, brown, red --- see legend), and 
the horizontal lines are the dynamical (green dashed) and escape (light green solid) timescales. 
 \label{3C279_timescales}}
\end{figure}

The shock acceleration-generated thermal + non-thermal electron spectra serve as a particle injection term into
simulations of subsequent radiative cooling of the electrons. To keep the number of 
parameter variations to a minimum, as in BBS17, here we will adopt a shock speed of 
\teq{v_{\rm s}=0.71c}, a magnetic field obliquity to the shock normal of  \teq{\ThetaBfone =32.3^{\circ}}, 
an upstream gas temperature of \teq{5.45 \times 10^7}K, and a velocity compression
ratio of \teq{r=3.71}.  
These choices well represent the environment of a strong, subluminal, 
mildly-relativistic shock.  \cite{SB12} note that there is modest sensitivity of the 
accelerated electron distributions to the magnetic obliquity \teq{\ThetaBfone}, and 
also that changes in the electron temperature will alter the velocity compression ratio 
across the shock, and the distributions somewhat.  Yet the objective of this paper 
is to identify the key acceleration characteristics that are required to successfully
model time-dependent, MW blazar spectra in a two-zone construct.  Accordingly, 
focusing on a fairly representative shock set up suffices for these goals, and 
an extensive exploration of spectral model variations with shock parameters 
is deferred to future work. 

As relevant radiative mechanisms, 
synchrotron radiation in a tangled magnetic field, synchrotron self-Compton (SSC) radiation, and 
inverse Compton scattering of external radiation fields (external inverse Compton = EIC) on various plausible target 
photon fields are taken into account in our simulations. All the relevant cooling rates, 
emissivities, and absorption coefficients are evaluated using the routines
described in detail in \cite{Boettcher13}. Particles may also leave the emission region 
on a time scale parameterized as a multiple of the light-crossing time scale of the emission region, 
$t_{\rm esc, e} = \eta_{\rm esc} \, \ell_{\rm rad} / c$. 
Thus, \teq{\sqrt{\eta_{\rm esc}}\, \ell_{\rm rad}} approximately represents 
(for \teq{\eta_{\rm esc}\gg 1}) the diffusive 
mean free path in long wavelength MHD turbulence in the radiation zone, 
with values far exceeding the short, gyro-scale 
pathlengths encountered by low energy charges undergoing DSA within the 
confines of the very turbulent shock layer.

Figure \ref{3C279_timescales} shows the energy 
dependence of the relevant time scales for the steady state generated to describe the quiescent-state 
multi-wavelength emission of 3C~279 (see Section \ref{sec:3C279}). DSA+SDA will be effective up to an energy 
$\gammax$, where the radiative cooling time scale becomes shorter than the acceleration time 
scale. Figure \ref{3C279_timescales} illustrates that for almost all particles at lower energies,
$\gamma < \gammax$, the acceleration time scale is many orders of magnitude shorter than the
radiative cooling and/or escape timescales. This implies that for particles of all energies significantly
below $\gammax$, the DSA process acts effectively instantaneously, while radiative cooling or escape
are negligible. Thus, numerically, DSA may be well represented as an instantaneous injection of relativistic 
particles at a (time-dependent) rate \teq{Q_e (\gamma_e, t)} [cm$^{-3}$ s$^{-1}$], which is then followed by 
evolution on the radiative and escape time scales in the larger emission zone. 
For our 3C 279 case study, \teq{\gammax \sim (2 - 3) \times 10^3}.  For Mrk 501, 
it is substantially higher at \teq{\gammax \sim 4 \times 10^5}, similar to the values derived in BBS17.

The injection function \teq{Q_e (\gamma_e, t)} is the distribution computed in the MC simulation, folded with 
an exponential cutoff of the form \teq{\exp ( - \gamma /\gammax ) }.
The normalization of the injection function \teq{Q_e (\gamma_e, t)} is determined through an injection luminosity, 
\teq{L_{\rm inj}} (in the co-moving jet frame), as
\begin{equation}
   L_{\rm inj} = {4 \pi \over 3} \, \ell_{\rm rad}^3 \, m_e c^2 \, \int\limits_1^{\infty} Q_e (\gamma_e, t) 
   \, \gamma_e \, d\gamma_e \quad .
 \label{eq:Linj}
\end{equation}
A simplification adopted in this first exploration of time-dependent radiation signatures
of relativistic shocks in AGN jets, is that we assume that the shock conditions and diffusion parameters
($\eta_1$, $\alpha$) remain constant during the passage of the shock (injection), and separately constant also 
during the quiescent phases before and after the shock passage.  Accordingly $Q_{e} (\gamma, t)$ 
involves a simple Heaviside step function in time, with the shock passage lasting 
a mere few hours.  This simplification of the shock evolution is a first approximation, noting that
the only observational constraints on changing shock conditions are based on the flux variability on
the observed variability timescale, and this timescale is captured in our model prescription. We defer the study of 
a more realistic and self-consistent time dependence of pitch-angle diffusion parameters and the 
resulting $Q_{e} (\gamma, t)$ to future work.

The distribution of relativistic electrons is assumed to be isotropic in the co-moving frame of the emission
region, and its evolution is simulated by numerically solving a Fokker-Planck equation of the form
\begin{equation}
    {\partial n_e (\gamma_e, t) \over \partial t} = - {\partial \over \partial \gamma_e} \Bigl( \dot\gamma_e
    \, n_e [\gamma_e, t] \Bigr) - {n_e (\gamma_e, t) \over t_{\rm esc, e}} + Q_e (\gamma_e, t) \, .
  \label{eq:FP}
\end{equation}
The solution is obtained using an implicit Crank-Nicholson scheme as described in \cite{BC02}. In Eq. (\ref{eq:FP}),
$\dot\gamma_e$ represents the combined radiative energy loss rate of the electrons, and all quantities are in 
the co-moving frame of the emission region, and the electron escape time scale is parameterized as
a multiple of the light-crossing time scale, $t_{\rm esc, e} = \eta_{\rm esc} \, R/c$.  

Radiation transfer is handled by forward evolution of a continuity equation for the photons,
\begin{equation}
    {\partial n_{\rm ph} (\epsilon, t) \over \partial t} = {4 \, \pi \, j_{\epsilon} \over \epsilon \, m_e c^2} 
    - c \, \kappa_{\epsilon} \, n_{\rm ph} (\epsilon, t) - {n_{\rm ph} (\epsilon, t) \over t_{\rm esc, ph}} \, .
  \label{eq:radiation}
\end{equation}
Here $j_{\epsilon}$ and $\kappa_{\epsilon}$ are the emissivity and absorption coefficient, respectively,
$\epsilon = h \nu / (m_e c^2)$ is the dimensionless photon energy, and $t_{\rm esc, ph}$ is the photon escape
time scale, $t_{\rm esc, ph} = (4/3) \, \ell_{\rm rad} / c$ for a spherical geometry \citep{Boettcher97}. 
Because of the tangled magnetic field assumed in the radiation zone, the synchrotron photons 
that seed the SSC signal are presumed isotropic.  In contrast, the external radiation field that is 
upscattered to form the EIC component is
assumed to be isotropic in the AGN rest frame, as appropriate for the BLR or dust torus radiation fields, as long 
as these seed photons are emitted within the BLR radius or the dust torus, respectively.  
Accordingly, this field is Doppler-boosted and highly anisotropic in the jet frame, thereby 
strongly enhancing the EIC emissivity. The total observed flux from the synchrotron, 
SSC and EIC emission is provided by the escaping photons, such that
\begin{equation}
    \nu F_{\nu}^{\rm obs} (\nu_{\rm obs}, t_{\rm obs}) = {\epsilon^2 \, m_e c^2 \, n_{\rm ph} (\epsilon, t) \,
    \delta^4 \, V_{\rm rad} \over 4 \pi \, d_L^2 \, (1 + z) \, t_{\rm esc, ph}} \quad ,
  \label{eq:nuFnu}
\end{equation}
where $\epsilon = (1 + z) \epsilon_{\rm obs} / \delta$ and $V_{\rm rad} \approx (4/3) \, \pi \, \ell_{\rm rad}^3$
is the co-moving volume of the emission region.  The jet-frame and observer time intervals are related through 
$\Delta t_{\rm obs} = \Delta t \, (1 + z) / \delta$. Our code outputs snap-shot SEDs and multi-wavelength light 
curves at 7 pre-specified frequencies $\nu_{i}$. For the present work, we chose $\nu_i$ as listed in Table 
\ref{LCfrequencies}.
All radiation spectra are corrected for $\gamma\gamma$ absorption by the Extragalactic Background
Light (EBL) using the model of \cite{Finke10}. However, for the test cases discussed below, 
the effect of EBL absorption is small (particularly for 3C 279 with little flare emission above 10 GeV)
and has no effect on the resulting light curve or cross-correlation features.

\begin{table}[ht]
\caption{\label{LCfrequencies}Observed frequencies for which light curves and local spectral indices are 
extracted in our simulations. }
\smallskip
\begin{center}
\begin{tabular}{cccc}
\hline
\noalign{\smallskip}
No. & Frequency & Band & Blazar\\
\noalign{\smallskip}
\hline
\noalign{\smallskip}
$\nu_1^{\rm obs}$ & 230 GHz & Radio & 3C 279\\
$\nu_2^{\rm obs}$ & $5.5 \times 10^{14}$~Hz & Optical R-band & 3C 279, Mrk 501\\
$\nu_3^{\rm obs}$ & $2.4 \times 10^{17}$~Hz & 1 keV X-rays & 3C 279, Mrk 501 \\
$\nu_4^{\rm obs}$ & $2.4 \times 10^{18}$~Hz & 10 keV X-rays & Mrk 501 \\
$\nu_5^{\rm obs}$ & $2.4 \times 10^{19}$~Hz & 100 keV X-rays & Mrk 501 \\
$\nu_6^{\rm obs}$ & $2.4 \times 10^{23}$~Hz & 1 GeV $\gamma$-rays & 3C 279, Mrk 501 \\
$\nu_7^{\rm obs}$ & $2.4 \times 10^{26}$~Hz & 1 TeV $\gamma$-rays & 3C 279, Mrk 501 \\
\end{tabular}
\end{center}
\end{table}

For the purpose of producing hardness-intensity diagrams, our code also extracts local spectral indices at 
the frequencies $\nu_i$ for each time step. Correlations between the light curves at different frequencies 
and possible inter-band time lags \teq{\tau} are evaluated using the Discrete Correlation Function 
(DCF) analysis as detailed in \citep{EK88}. This is a discretization of the correlation function
\begin{equation}
     {\cal C}_{a,b}(\tau )\; \equiv\; \dover{1}{{\cal N}_a{\cal N}_b}
    \displaystyle\int_{-T}^T F^{\rm obs}_a(t)\, F^{\rm obs}_b(\tau -t ) \, dt
 \label{eq:DCF_def}
\end{equation}
which accounts for errors due to uneven sampling. Here, fluxes $F_{a,b}^{\rm obs}$ are in wavebands a,b, 
with
\begin{equation}
   {\cal N}_a \; =\; \int_{-T}^T  F^{\rm obs}_a(t)\, dt
 \label{eq:DCF_norm}
\end{equation}
defining the normalizations, and the times $\tau$ and $t$ being implicitly 
in the observer frame.  The bracketing time \teq{T} is chosen large enough that 
the solutions realize the long-term quiescent state.  
As will become evident in due course, 
the lags \teq{\tau} will be tightly coupled to the relative cooling times in 
different wavebands.

For each flare simulation, we first let the radiation code run until it reaches a stable equilibrium with a set of
quiescent-state parameters. An individual flaring event is then simulated by changing various input parameters 
as a function of time. The default mode for such changes will be a step function in time for the duration 
$\Delta t = \ell_{\rm rad} / v_{\rm s}$ in the co-moving frame of the emission region. The value of the 
shock speed \teq{v_{\rm s}} in the jet frame is that used in the Monte-Carlo acceleration code to  simulate
the electron injection spectra; this is fixed at the representative value of \teq{v_{\rm s} = 0.71c}.

\section{Results}
 \label{sec:results}

The code described in the previous section has been applied to two test cases: (1) Two multi-wavelength flares
of the FSRQ 3C~279 during the active period in 2013 - 2014 \citep{Hayashida15}, 
and (2) a generic test case for the typical SED and 
variability patterns of the prototypical HBL Mrk 501 \citep[e.g.,][]{Abdo11,Ahnen18}.  
These present contrasting examples that evince a range of spectral and temporal character.

\subsection{Application to 3C~279}
 \label{sec:3C279}

The FSRQ 3C~279, located at a redshift of $z = 0.536$ \citep{Lynds65}, gained prominence due to its exceptional
gamma-ray activity during the early days of the Energetic Gamma-Ray Experiment Telescope (EGRET) on board the 
{\it Compton} Gamma-Ray Observatory (CGRO) in the early 1990s \citep[e.g,][]{Wehrle98,Hartman01}. It continues to 
be one of the brightest gamma-ray blazars detected by the Large Area Telescope (LAT) on board the {\it Fermi}
Gamma-Ray Space Telescope \citep[e.g.,][]{Abdo10}, and is one of only a handful of FSRQs also detected in
very-high-energy (VHE: $E > 100$~GeV) gamma rays by ground-based Imaging Atmospheric \v{C}erenkov Telescopes
\citep[IACTs; e.g.,][]{Teshima08,DeNaurois2018}. 

\begin{figure}[ht]
\vspace{0.5cm}
\centering
\centerline{\includegraphics[width=8cm]{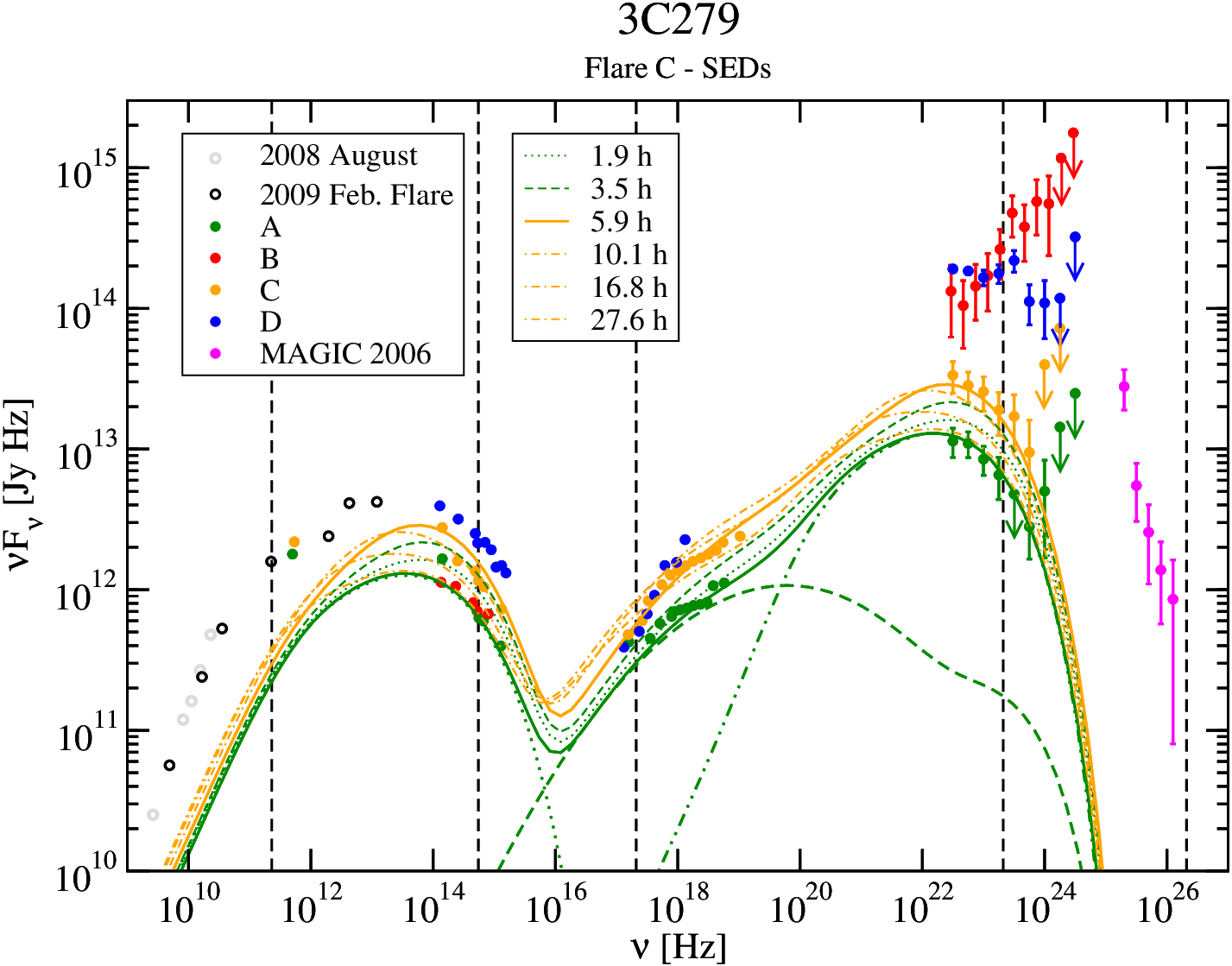}}
\caption{Snap-shot SEDs of 3C~279 during 2013 -- 2014, with a model simulation to reproduce Flare C. Data are 
from \cite{Hayashida15}. The heavy solid green curves show the quiescent-state (period A) fit, with individual 
radiation components shown as dotted (synchrotron), dashed (SSC) and dot-dot-dashed (EIC on dust torus photons) 
curves. Light green curves illustrate the spectral evolution during the rising part of the simulated Flare C; 
yellow curves show the evolution during the decaying part. The dashed vertical lines indicate the frequencies
at which light curves and hardness-intensity diagrams are extracted (see Table~1). }
 \label{3C279_SEDs_FlareC}
\end{figure} 

Extensive multi-wavelength observations of 3C~279 during flaring activity in the period 
December 2013 -- April 2014 were reported in
\cite{Hayashida15}. Figure 7 of that paper shows multi-wavelength light curves of 3C~279, where several
$\gamma$-ray flares (B, C, D) are identified, in addition to a quiescent period (A). Figure \ref{3C279_SEDs_FlareC}
shows snap-shot SEDs extracted by \cite{Hayashida15} for these episodes, along with our model simulation to 
reproduce Flare C. The quiescent state (period A) has been reproduced with the model parameters listed in
Table \ref{3C279parameters}. The characteristic time scale of short-term flares of 3C~279 during the 2013 -- 2014 period 
(including Flare C) is $\Delta t^{\rm obs} \sim 0.3 - 1$~day.  Throughout this paper, 
we assume a viewing angle of $\theta^{\ast}_{\rm obs} \approx 1/\Gamma^{\ast}$, and a typical Doppler factor 
of $\delta \approx \Gamma^{\ast} = 15$, 
a value that satisfies pair compactness lower bounds of 
\teq{\delta \gtrsim 8-10} for 3C 279 as discussed in \cite{Maraschi92ApJ} and \cite{Ghisellini93ApJ}.  Then, for
this \teq{\delta} and given the redshift of $z = 0.536$, for a mildly relativistic shock with \teq{v_{\rm s} \sim 0.7c}, 
the variability timescale implies a size of the active region of 
$\ell_{\rm rad} \sim 1.8 \times 10^{16}$~cm.

\begin{table}[ht]
\caption{\label{3C279parameters}Model parameters for the fit to 3C-279 SEDs during the quiescent-state (period A).}
\smallskip
\begin{center}
\begin{tabular}{cc}
\hline
\noalign{\smallskip}
Parameter & Value \\
\noalign{\smallskip}
\hline
\noalign{\smallskip}
\multicolumn{2}{c}{Jet-frame parameters} \\
\noalign{\smallskip}
\hline
\noalign{\smallskip}
Electron injection luminosity & $L_{\rm inj, q} = 1.1 \times 10^{43}$~erg~s$^{-1}$ \\
Emission region size & $\ell_{\rm rad} = 1.8 \times 10^{16}$~cm \\
Jet-frame magnetic field & $B = 0.65$~G \\
Escape time scale parameter & $\eta_{\rm esc} = 3$ \\
Thermal $e^+e^-$ density & $n_e = 1.2 \times 10^4$ cm${}^{-3}$ \\
PAS m.f.p. low-energy limit & $\eta_1 = 100$ \\
PAS m.f.p. scaling index & $\alpha = 3$ \\
\noalign{\smallskip}
\hline
\noalign{\smallskip}
\multicolumn{2}{c}{AGN-frame parameters} \\
\noalign{\smallskip}
\hline
\noalign{\smallskip}
Bulk Lorentz factor & $\Gamma^{\ast} = \delta = 15$ \\
Accretion-disk luminosity & $L^{\ast}_d = 6 \times 10^{45}$~erg~s$^{-1}$ \\
Distance from BH & $z^{\ast}_i = 0.1$~pc \\
Ext. rad. field energy density & $u^{\ast}_{\rm ext} = 4 \times 10^{-4}$~erg~cm$^{-3}$ \\
Ext. rad. field BB temperature & $T^{\ast}_{\rm ext} = 300$~K. \\
\end{tabular}
\end{center}
\end{table}

We find that, to model the quiescent-state SED of 3C~279, EIC needs to be dominated by scattering of a
low-temperature external radiation field, as expected to arise from the dusty torus. For the present 
study, we approximate it as a thermal blackbody at a temperature of $T^{\ast}_{\rm ext} = 300$~K. 
The quiescent state fit is illustrated by the solid green line in Fig. \ref{3C279_SEDs_FlareC}. We find 
that it can be well described with an electron injection spectrum produced by DSA+SDA with a pitch-angle 
scattering mean free path scaling as $\lambda_{\rm pas} = 100 \, r_g \, (p/m_ec)^2$, i.e., $\lambda_{\rm pas} 
\propto p^3$. Based on the competition of acceleration and cooling time scales, as illustrated in 
Fig.~\ref{3C279_timescales}, electrons are accelerated up to a maximum energy of $\gammax = 
2.4 \times 10^3$. In the quiescent-state equilibrium, the magnetization of the emission region is 
$u_B/u_e = 0.094$, which satisfies the requirement of a weakly magnetized medium for the 
formation of a strong shock. Note that $u_B/u_e$ relates to the non-relativistic magnetization
$\Sigma$ as listed in Table~\ref{3C279_flare_parameters} via
$u_B/u_e = \Sigma/(2 \langle\gamma_e\rangle)$, where $\langle\gamma_e\rangle$ is
the average (relativistic) electron Lorentz factor.

We point out that many of the emission-region parameter values are degenerate in the sense that 
they depend on the assumed value of the bulk Lorentz factor $\Gamma^{\ast}$ and the Doppler factor $\delta$,
which have been assigned typical values for this source. However, within reasonable bounds on $\Gamma^{\ast}$
and $\delta$, the general conclusions concerning the plasma physics and turbulence characteristics will 
not change. In particular, we emphasize that the location of the synchrotron peak (modulo the Doppler factor) 
closely constrains the value of $\lambda_{\rm pas} (\gammax)$, as it is independent of the magnetic field. 
This leaves a degeneracy between $\eta_1$ and $\alpha$, which would change primarily the thermal-to-non-thermal 
particle density ratio, but not significantly alter the radiative signatures. As in BBS17, 
there is an approximate tolerance of about \teq{\pm 0.2} in \teq{\alpha}, and a tolerance of a factor 
of \teq{\sim 1-5} in \teq{\eta_1} permitting MW spectral fits of similar character and precision.

\subsubsection{3C~279 --- Flare C} 
 \label{sec:flareC}

For the study of expected multi-wavelength variability in the internal-shock model, we first focus 
on Flare C (yellow in Figure \ref{3C279_SEDs_FlareC}), which is characterized by an almost unchanged
Compton dominance compared to the quiescent state during period A.  One natural interpretation 
is that this and other flares closely sample the accelerator/injector \citep[e.g.,][]{Yan16mnras}.
Therefore, such flaring behaviour can plausibly be reproduced by merely increasing the number of radiating 
non-thermal electrons generated by the shock, thus enhancing the synchrotron 
and EIC emission at the same rate. Specifically, in our simulation, after reaching a steady state with
the input parameters listed in Table \ref{3C279parameters}, we increase the electron injection luminosity 
to $L_{\rm inj, f} = 5.0 \times 10^{43}$~erg~s$^{-1}$, i.e., about 4.5 times its value  $L_{\rm inj, q}$ during the 
quiescent state. All parameter changes for the flaring episodes C and B (see Section \ref{sec:flareB}) are 
summarized in Table \ref{3C279_flare_parameters}.  To identify more intimately the changes
in physical conditions in the jet, those listed include the derived parameters of the cyclotron 
frequency, \teq{\ecyc}, the thermal electron number density $n_e$ at the end of the flare injection episode, 
the plasma frequency, \teq{\wp} and the non-relativistic magnetization,
\teq{\Sigma = B^2/[\, 4\pi n_e m_ec^2\, ] \equiv \ecyc^2/\wp^2}.

\begin{table}[ht]
\caption{\label{3C279_flare_parameters}Parameters adopted to reproduce the quiescent state 
and flares C and B of 3C~279. All values are in the jet frame. }
\vspace{-12pt}
\begin{center}
\begin{tabular}{cccc}
\hline
\noalign{\smallskip}
Parameter${}^{\dag}$ & Quiescent & Flare C & Flare B \\
&& Dec. 31, 2013 & Dec. 20, 2013 \\
\noalign{\smallskip}
\hline
\noalign{\smallskip}
$L_{\rm inj}$ [erg~s$^{-1}$] & $1.1 \times 10^{43}$ & $5.0 \times 10^{43}$ & $4.0 \times 10^{44}$ \\
$B$ [G] & 0.65 & 0.65 & 0.075$^\ast$ \\ 
$\eta_1$ & 100 & 100 & 10 \\
$\alpha$ & 3 & 3 & 2.3 \\[2pt]
\hline
$n_e$ [cm${}^{-3}$] & $1.2 \times 10^4$ & $2.23 \times 10^4$ & $2.15 \times 10^4$ \\
$\ecyc$ [MHz] & 11.4 & 11.4 & 1.32 \\[2pt]
$\wp$ [MHz] & 6.18 & 8.42 & 8.27 \\[2pt]
$\Sigma = \ecyc^2/\wp^2$ & 3.42 & 1.84 & 0.025 \\[2pt]
\hline
\end{tabular}
\end{center}
\vspace{5pt}
{\bf Note}: 
${}^{\dag}$ The cyclotron frequency $\ecyc$, the electron number density $n_e$, 
the plasma frequency $\wp$ and magnetization $\Sigma$ are values derived using $B$ and $L_{\rm inj}$.
$^\ast$This field is at the outset of an exponential recovery,
described in Eq.~(\ref{eq:Bfield_recovery}), i.e. \teq{B_f} therein.
\end{table}

\begin{figure}[ht]
\vspace{0.5cm}
\centering
\includegraphics[width=8cm]{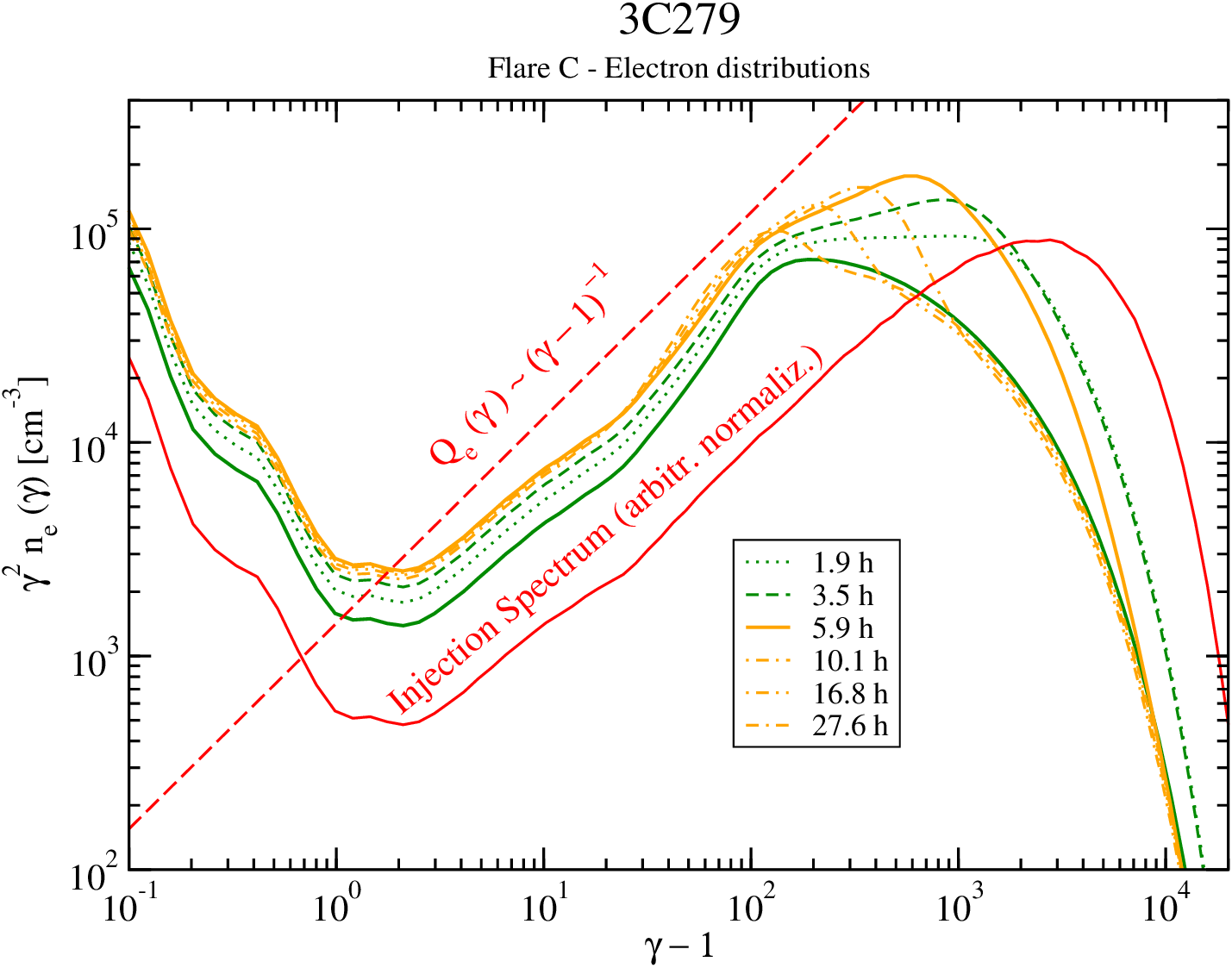}
\caption{Electron distribution sequence \teq{n_e(\gamma_e, \, t)} 
corresponding to the Flare C simulation illustrated in Fig.~\ref{3C279_SEDs_FlareC}. 
The diagonal line marks the approximate shock injection power law distribution that 
results primarily from the shock drift (SDA) mechanism: see \cite{SB12}, BBS17.}
 \label{fig:FlareC_edist}
\end{figure}   

Snap-shot SEDs during various times of the spectral evolution of this Flare C simulation 
are shown by the light curves in Figure \ref{3C279_SEDs_FlareC}, with light green curves illustrating 
the rising portion, and light yellow curves illustrating the decaying phase of the flare. The heavy yellow 
curve shows the SED during the peak of the flare. The excellent fit to both the period A and C SEDs
indicates that such variability can be produced without any changes of the turbulence and particle 
acceleration characteristics, as long as a 4.5-fold increase in the injection 
luminosity \teq{L_{\rm inj}} of accelerated electrons
can be achieved.  We note that this amplification factor differs from the factor of 1.86 enhancement 
in the density $n_e$ of the thermal electron population (see Table~\ref{3C279_flare_parameters}), 
because of the influences of cooling 
and escape on the electron distribution function: see Eq.~(\ref{eq:FP}).
The evolution of the electron distribution throughout the Flare C sequence 
is depicted in Figure~\ref{fig:FlareC_edist}, clearly illustrating the competition between 
acceleration and cooling at the highest Lorentz factors.

\begin{figure}[ht]
\vspace{0.5cm}
\centering
\includegraphics[width=8cm]{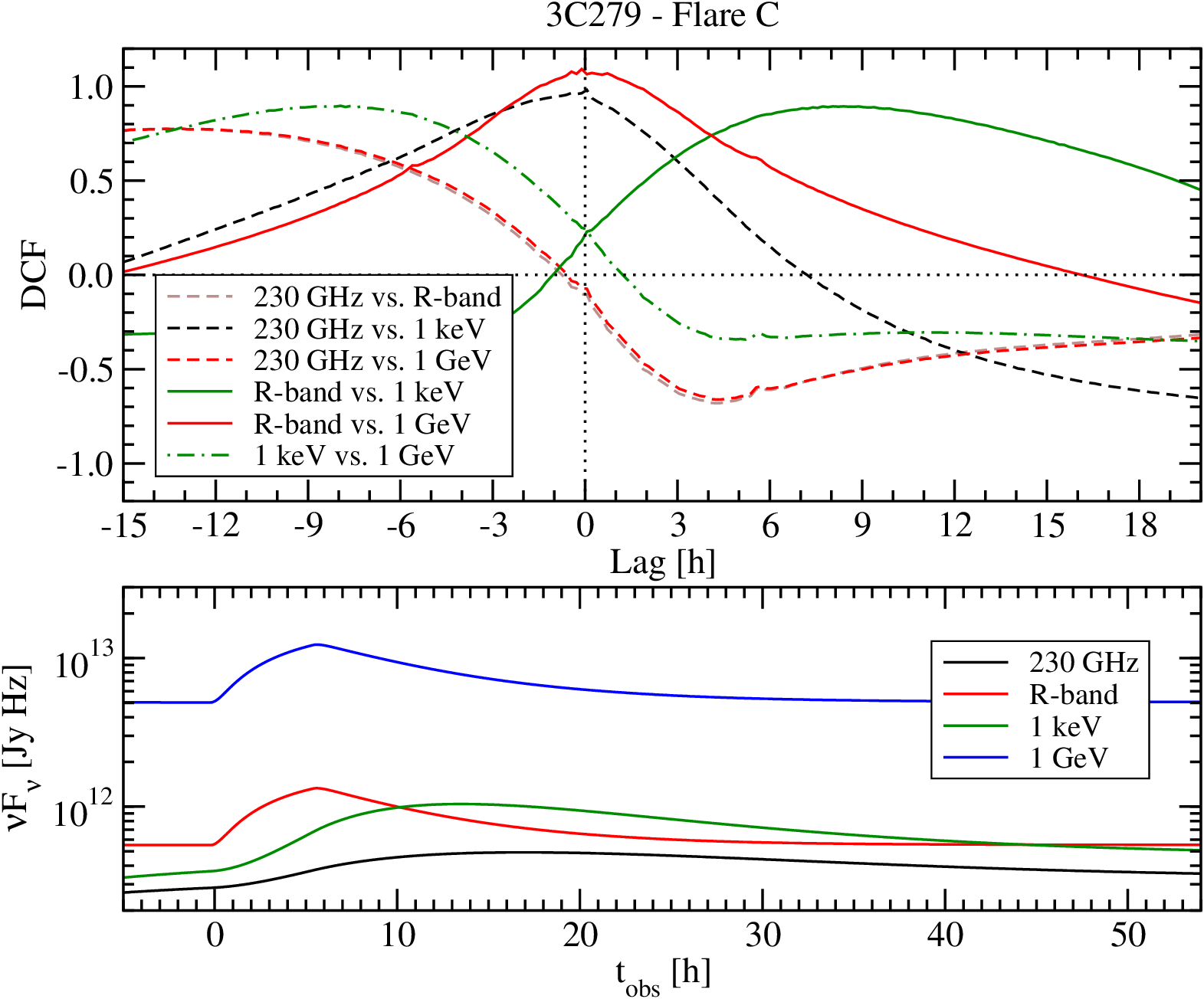}
\caption{{\it Bottom:} Multi-wavelength light curves extracted from the Flare C simulation illustrated in 
Fig. \ref{3C279_SEDs_FlareC}. Injection from shock acceleration starts at \teq{t_{\rm obs}=0} and ends 
at \teq{t_{\rm obs}=5.5}hr, after which cooling reduces the R Band and GeV fluxes.
{\it Top:} Discrete cross correlation functions evaluated from the light curves shown in the bottom panel: 
see Eq.~(\ref{eq:DCF_def}) and associated text for details.}
\label{LC_FlareC}
\end{figure}

The resulting light curves in the {\sl mm }radio, optical, X-ray and GeV $\gamma$-ray bands are 
illustrated in the bottom panel of Fig. \ref{LC_FlareC}. Unfortunately, the observational data for 
Flare C are too sparsely sampled to warrant a detailed comparison between observations and the model 
light curves.
No significant TeV emission is predicted by our simulation, in accordance
with the finding by \cite{Boettcher09} that leptonic models have difficulties reproducing the VHE emission 
observed in several exceptional flare states of 3C~279.  The physical origin of this 
is in the low value of \teq{\gammax}, imposed by the very strong Compton cooling in the emission region,
and required to generate the low synchrotron peak frequency.
No VHE emission was detected from 3C~279 during the 
2013 -- 2014 flaring episodes discussed in \cite{Hayashida15},
though there have been detections by IACTs, for example by MAGIC in 2006 
which is included as archival VHE flux points (magenta) 
in Figure~\ref{3C279_SEDs_FlareC}.  Using these light curves, 
cross-correlations between the various frequency bands can be calculated; these 
are depicted in the top panel of Fig.~\ref{LC_FlareC}; see also \cite{Teshima08}.

The model predicts, as expected in most leptonic single-emission-zone models for FSRQs, 
that the optical and $\gamma$-ray light 
curves are closely correlated with zero time lag, as those bands are produced by synchrotron and Compton emission 
from electrons of similar energies, sampling electrons with energies near \teq{\gammax}. They therefore possess 
comparable cooling times
that are short, driving the prompt declines in flux seen in Figure~\ref{LC_FlareC} once the injection is terminated.
The X-ray emission, being dominated by
SSC emission of low-energy electrons with significantly longer cooling time scales than those producing the optical 
and GeV $\gamma$-ray emission, is expected to lag behind the optical and $\gamma$-ray emissions by $\sim 8$~hr,
while the {\sl mm} radio band is expected to show an even longer delay behind optical and $\gamma$-rays, with 
slightly weaker correlation. These are manifested in the much slower drops in X-ray and radio fluxes 
once the shock injection is shut off, occurring on timescales of 15--30 hours.  Unfortunately, the light curve coverage in 
most existing data for 3C 279 is not sufficient for a detailed, direct comparison of our predictions with observations. 
This includes the radio, optical and X-ray data reported in \cite{Hayashida15}.   Yet we observe that our model 
duration of around 15 hours for the GeV flare signal is very consistent with the duration of Flare C as observed 
by {\it Fermi}-LAT.  This, combined with the satisfactory spectral reproduction of MW data, instills a confidence in the robustness of the hybrid acceleration-emission modeling approach adopted here.

\begin{figure}[ht]
\vspace{0.5cm}
\centering
\includegraphics[width=8cm]{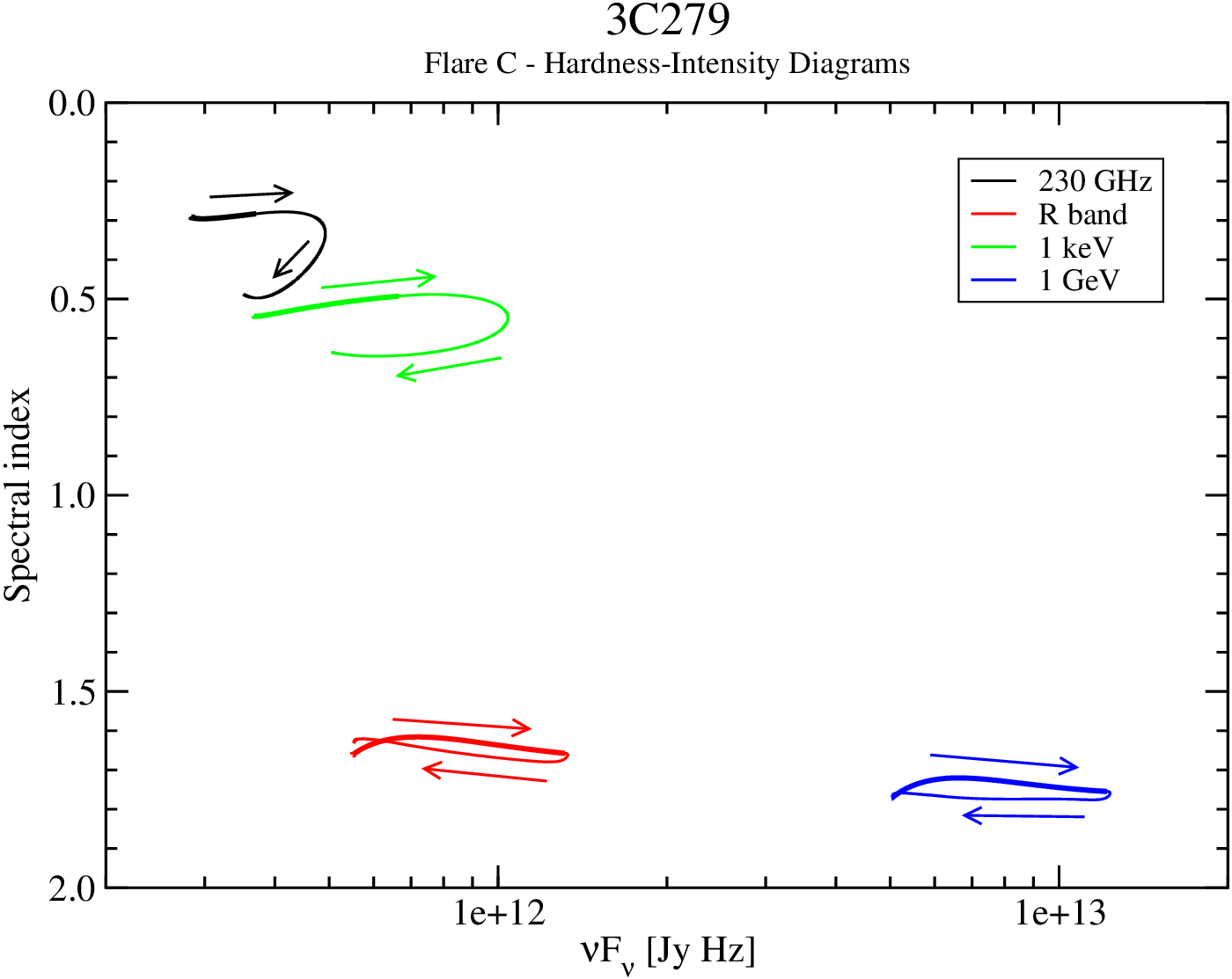}
\caption{Hardness-intensity diagrams extracted from the Flare C simulation illustrated in Figs. \ref{3C279_SEDs_FlareC}
--- \ref{LC_FlareC}, with the spectral index serving as a proxy for traditional hardness ratios. }
\label{FlareC_HIDs}
\end{figure}

Figure \ref{FlareC_HIDs}
shows the hardness-intensity diagrams (HIDs) extracted from our simulatons of flare C. 
The hardness is represented by the spectral index here and for all HIDs presented in the paper, with 
the index corresponding to the differential energy flux spectrum, i.e. \teq{F_{\nu}}.
While only very weak spectral variability is predicted in the optical and GeV $\gamma$-ray bands,
pronounced clockwise spectral hysteresis (harder rising-flux spectra; softer decaying-flux
spectra) is expected in the {\sl mm} radio and X-ray bands. Due to the typically faint X-ray fluxes from FSRQs
(X-rays covering the valley between the synchrotron and Compton spectral components),
it is difficult to discern such spectral hysteresis 
in the X-ray observations of such objects --- in contrast to HBLs, where the 
X-ray emission is synchrotron dominated \citep[e.g.,][]{Takahashi96}. Observing such features in other 
wavelength bands could enable stringent constraints on the magnetic field and the shock injection in the emission region 
\citep[see, e.g.,][]{Kirk98AA,Boettcher03}.  Yet the predicted spectral hysteresis at optical and GeV energies
may be too subtle to be detected by current-generation instrumentation.  However, we note that 
on longer time scales of \teq{\sim 25-30}days associated with general source variability, 
the WEBT campaign for 3C 279 detailed in \cite{Boettcher07ApJ} indicates for one time interval a counter-clockwise hysteresis 
loop in an optical B-R color versus R magnitude HID diagram in Fig.~5 therein, and for a preceding interval, a tilted 
figure 8 hysteresis profile.  Since these are
not closely related to shock-instigated flare activity, they provide no insights into our present models.

\subsubsection{3C279 --- Flare B}
 \label{sec:flareB}

\begin{figure}[ht]
\vspace{0.5cm}
\centering
\centerline{\includegraphics[width=8cm]{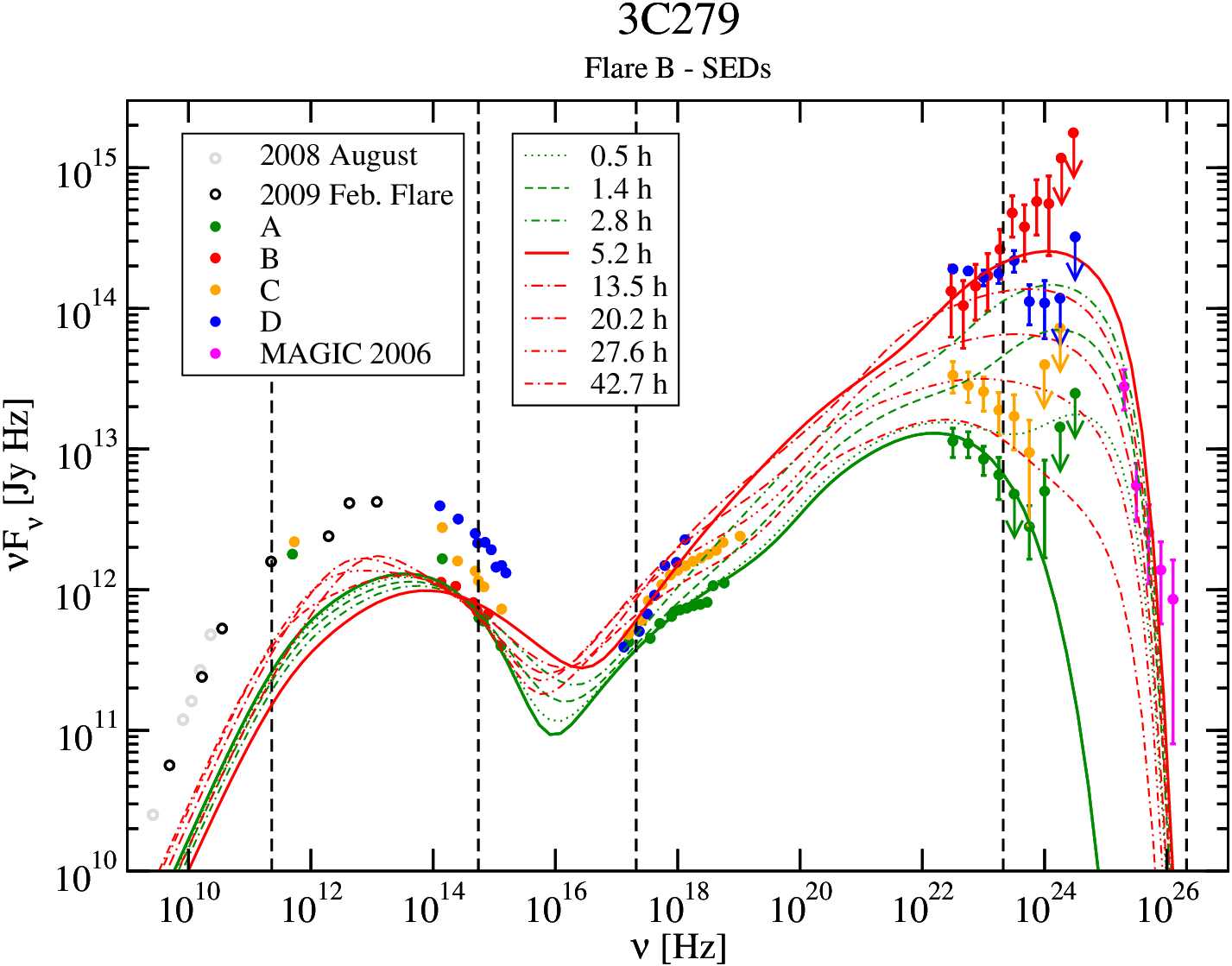}}
\caption{Snap-shot SEDs from the modeling of flare B of 3C~279 in December 2013 \citep[Data from ][]{Hayashida15}. 
The heavy solid green curves show the quiescent-state (period A) fit. Light green curves illustrate the spectral 
evolution during the rising part of the simulated Flare B; red curves show the evolution during the decaying part. 
The dashed vertical lines indicate the frequencies at which light curves and hardness-intensity diagrams are extracted.}
\label{3C279_SEDs_FlareB}
\end{figure}

Flare C discussed above is somewhat atypical for FSRQ flares as it exhibits a negligible increase 
in the Compton dominance compared to the quiescent state. A much more common occurrence are variability
patterns exhibiting larger flux ranges at higher energies, i.e., strongly increasing Compton
dominance during multi-wavelength flares, as is evident during flare B (red SEDs in Figs. \ref{3C279_SEDs_FlareC}
and \ref{3C279_SEDs_FlareB}). Such flaring behavior can plausibly be explained by a temporary increase of
the energy density of the external target photon field for EIC Compton scattering, e.g., in synchrotron
mirror scenarios as proposed by \cite{BD98,Tavani15,McDonald15,McDonald17}, which might not require any 
changes of the particle acceleration process in the emission region. The exploration of such scenarios is outside 
the scope of this paper, which is to study the effects of time-varying turbulence and particle acceleration 
characteristics in blazar jets. 

In this Subsection, we investigate what changes in magnetohydrodynamic turbulence, particle acceleration and other
source-intrinsic parameters would be required in order to reproduce flare B of \cite{Hayashida15} without 
invoking a temporary change of the external radiation field.  We choose flare B for this exercise, as it 
appears to present an especially challenging case of an ``orphan'' $\gamma$-ray flare with no significant 
counterpart in the optical (synchrotron) flux.

The $\gamma$-ray spectrum in the Flare B SED is significantly harder than the (episode A) quiescent SED. 
This requires significantly harder electron injection spectra. The simplest way to achieve this is via
a change of the turbulence/diffusion parameters $\eta_1 = 100 \to 10$ and $\alpha = 3 \to 2.3$. A good fit 
to the Flare B $\gamma$-ray SED could then be obtained if the electron injection
luminosity also changes to $L_{\rm inj, f} = 4.0 \times 10^{44}$~eg/s, i.e., $\sim 36$ times the quiescent
level. If this were the only change in parameters, the model would naturally predict an equally strong
flux flaring and spectral hardening in the synchrotron spectrum (especially in the optical), which is not observed. 
Within our single-radiation-zone model, keeping the optical flux constant during the $\gamma$-ray flare
requires a reduction of the magnetic field by an amount that exactly compensates for the increased 
injection of high-energy electrons. We find that with a change of $B = 0.65 \to 0.075$~G, the optical
spectrum will undergo only moderate spectral-index changes,
while maintaining its overall flux.  The SSC component is also not dramatically modified,
thereby avoiding an unobserved overproduction of X rays.

\begin{figure}[ht]
\vspace{0.5cm}
\centering
\includegraphics[width=8cm]{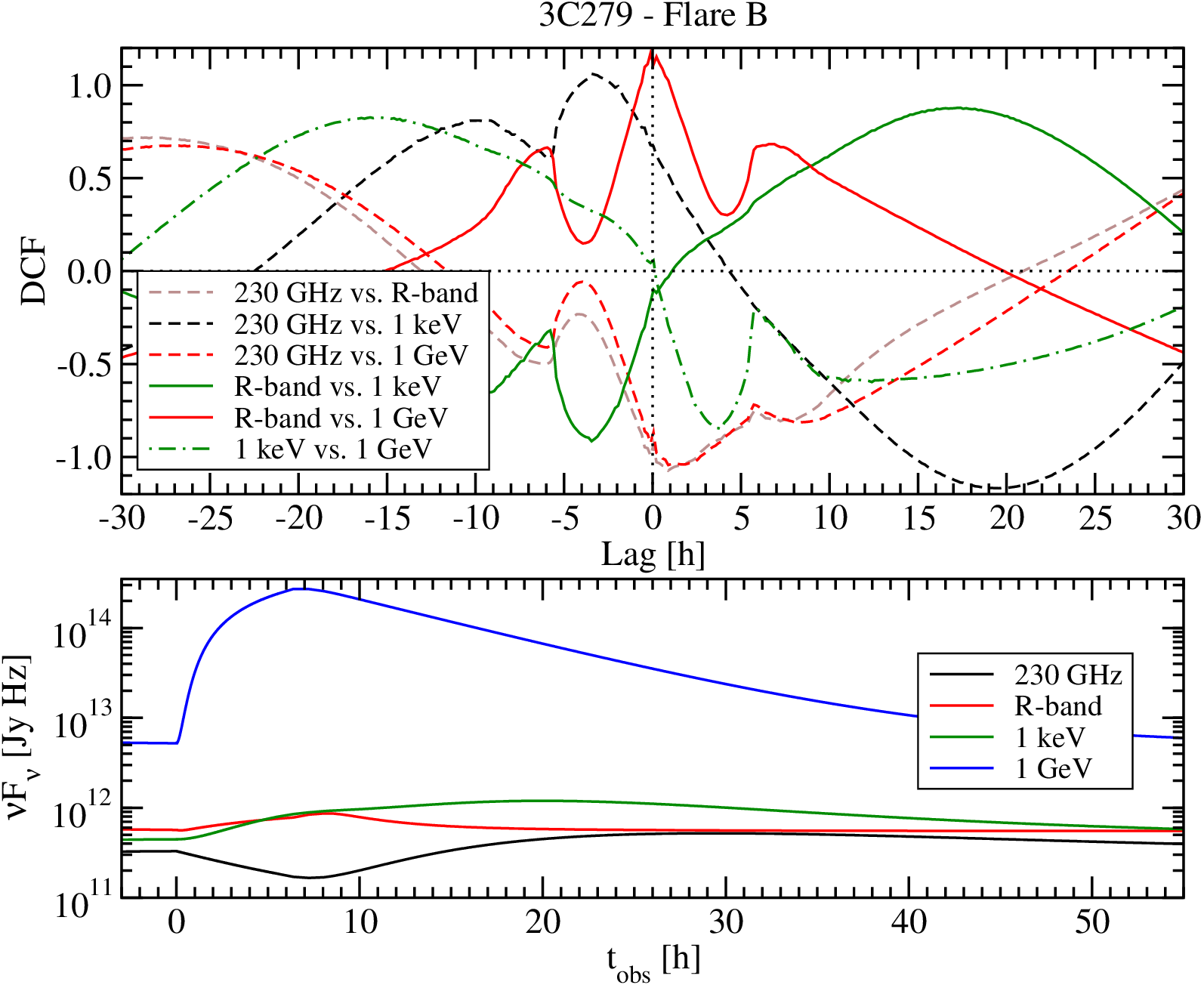}
\caption{{\it Bottom:} Multi-wavelength light curves extracted from the Flare B simulation illustrated in 
Fig. \ref{3C279_SEDs_FlareB}.
{\it Top:} Discrete cross correlation functions evaluated from the light curves shown in the bottom panel.}
\label{LC_FlareB}
\end{figure}   

Another issue occurs at the end of the flaring episode (i.e., when the shock causing the flare either leaves 
the emission region or loses its strength), where our standard model assumption was that all parameters revert to their 
quiescent-state values. At the end of the flare activation period, this additional injection
has built up a significant excess of high-energy electrons, from which it takes several days in the 
observer's frame to re-establish the quiescent-state electron distribution. Thus, if the magnetic 
field were to relax to its (higher) quiescent-state value immediately at the end of the flaring
injection episode, a large optical flare would result around $\sim 1$~day subsequent to the 
$\gamma$-ray Flare B, which has not been observed \citep{Hayashida15}. 
Suppressing this flare requires
that the magnetic field is only gradually restored to its quiescent-state value. A simulation that
does not predict significant optical variability could be achieved with a gradual restoration of 
the magnetic field of the form 
\begin{equation}
   B(t) = B_q + (B_f - B_q) \, e^{-(t' - t'_{\rm end}) / t'_{\rm rec}}
   \;\; ,\quad
   t' > t'_{\rm end} \; ,
 \label{eq:Bfield_recovery}
\end{equation}
i.e., after the end of the flare injection episode, $t'_{\rm end}$, on a time 
scale of $t'_{\rm rec} = 2.8 \times 10^5$~sec (in the co-moving frame). This time scale is of the order of 
the characteristic radiative cooling time scale of electrons that are emitting optical synchrotron photons
when 3C 279 is in its quiescent state, i.e. episode A. Here, \teq{B_f} is the field at the onset of Flare B, 
and \teq{B_q} is the long-term quiescent value as listed in Table~\ref{3C279_flare_parameters}.
In Section \ref{sec:discuss} we will discuss critically whether such 
a combined change of parameters could represent a realistic internal-shock scenario in a blazar jet.

Fig. \ref{LC_FlareB} shows, analogous to the case of flare C in the previous
sub-section, the multi-wavelength light curves and cross-correlation functions between different
wavelength bands. The combination of parameter changes differs from those for Flare C, and the gradual restoration 
of the magnetic field leads to more complicated variability patterns, especially in the optical,
as well as anti-correlated variability (a dip) in the radio light curve compared to all higher 
frequencies.  The origin of the dip in the radio flux is primarily the prompt reduction in the magnetic 
field at the onset of Flare B, which more than offsets the rise in the non-thermal electron density.
Interestingly, the 6-hour rise time of the 1 GeV light curve and its 10--15 hour $e$-fold 
decay timescale are fairly consistent with the high time-resolution {\it Fermi}-LAT data displayed 
in Fig.~2 of \cite{Hayashida15}.

\begin{figure}[ht]
\vspace{0.5cm}
\centering
\includegraphics[width=8cm]{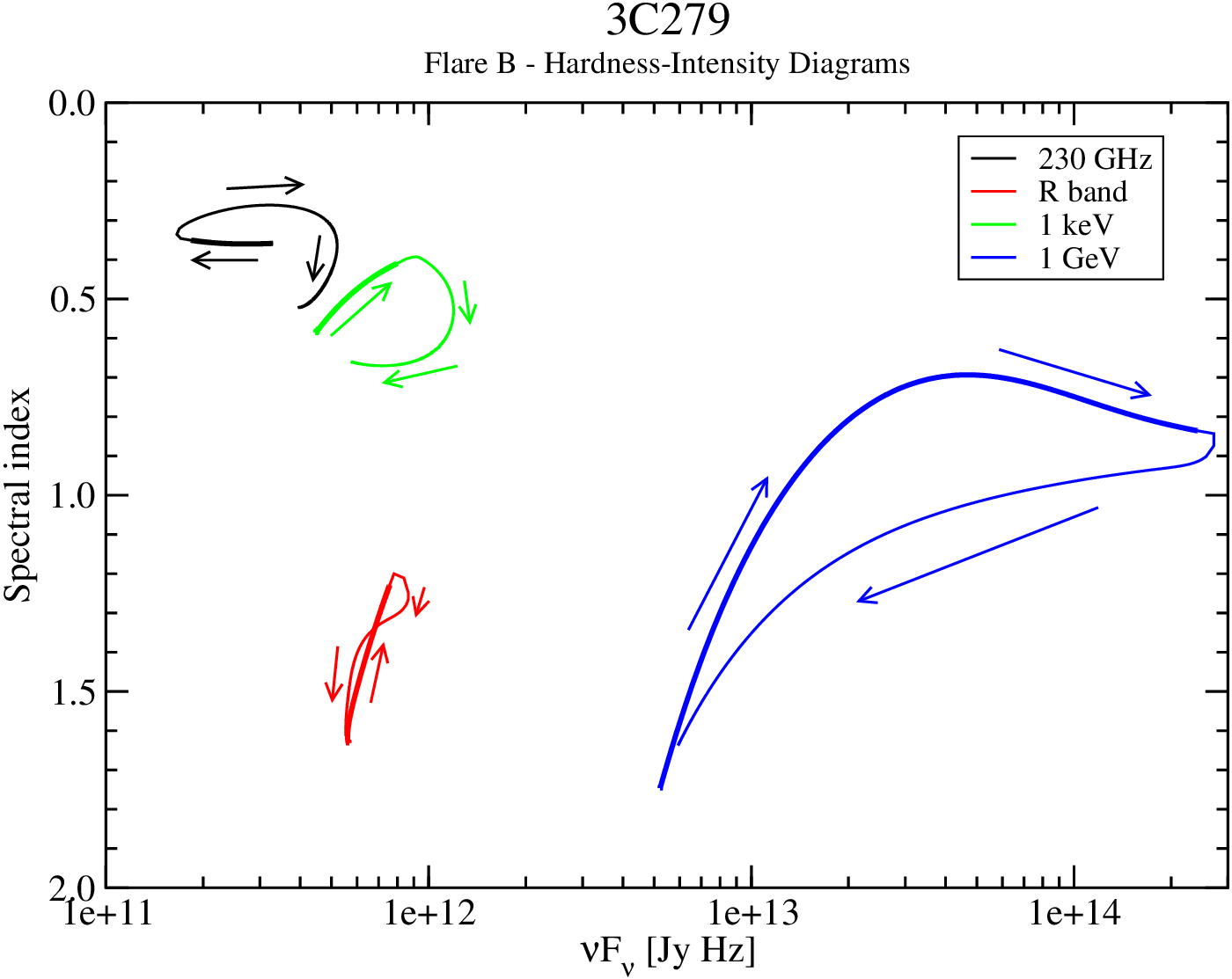}
\caption{Hardness-intensity diagrams extracted from the Flare B simulation illustrated in Figs. \ref{3C279_SEDs_FlareB}
--- \ref{LC_FlareB}. }
\label{FlareB_HIDs}
\end{figure}   

What little variability that remains in the optical light curve is expected to be well correlated 
with the $\gamma$-ray light curve, i.e. with zero delay. As in the case of flare C, the X-rays are 
delayed with respect to the $\gamma$-rays and optical emission, but with a significantly longer 
delay timescale of about $\sim 16$~hours. This is a direct consequence of the lower magnetic field and, 
thus, longer radiative cooling time scale of the low-energy electrons responsible for the SSC 
X-ray emission\footnote{Recall that the higher Compton dominance during flare B is achieved by 
a harder electron spectrum without a change of the external radiation field. Thus, a lower 
magnetic field will be directly reflected in a longer radiative cooling time scale.}. 

The hardness-intensity diagrams plotted in Fig. \ref{FlareB_HIDs} illustrate that the large-amplitude 
$\gamma$-ray flaring activity during flare B is predicted to be associated with significant clockwise
spectral hysteresis (as in flare C, harder spectrum during rising flux, softer during decaying flux), with 
spectral-index changes on $\sim $~a few hours time scales.  Such variations may be measurable with {\it Fermi}-LAT 
during the brightest $\gamma$-ray flares of 3C~279, as is evidenced by its temporal resolution 
of Flare B at the 3 hour level \citep[e.g., see Fig.~2 of ][]{Hayashida15}.  Significant spectral hysteresis, 
similar to that 
predicted for flare C, is also predicted in the X-rays, but their detection may be hampered by the 
relatively low X-ray flux of FSRQs like 3C~279. 

Table~\ref{3C279_flare_parameters} lists three derived parameters for Flares B and C that 
inform the physical process of acceleration.  The first of these is the 
electron gyrofrequency, \teq{\ecyc}, which represents the fundamental scale 
of the Fermi DSA process when it is efficient.  It is of the order of a few MHz for 
our quiescent emission and Flare B and C models.  In our implementation, the 
acceleration rate is \teq{d\gamma /dt \sim \ecyc/\eta (p)} with \teq{\eta (p)\gg 1} 
ensuring that the energization rate is much slower than Bohm-limited DSA and 
SDA drives the acceleration process.  The plasma frequency \teq{\wp} is of similar 
order to \teq{\ecyc} in all three models so that the magnetization \teq{\sigma = \ecyc^2/\wp^2} 
ranges between \teq{0.02} and 4.  Since \teq{c/\wp} is the inertial scale in electrodynamic
systems, \teq{\wp} defines the rate of Wiebel-instability driven 
acceleration in shocks imbued with strong MHD turbulence, or an approximate scaling 
for the magnetic reconnection acceleration rate in relativistic systems with multiple sites 
distributed amid converging magnetic islands -- see the discussion in BBS17.  
Therefore, \teq{\wp \gg \ecyc/\eta (p)} and from this one infers that acceleration 
driven by either reconnection or by the Weibel instability in their basic forms is too efficient to 
accommodate the infrared or optical synchrotron peak in 3C 279.  They essentially 
are too turbulent, as is the Bohm-limited Fermi shock configuration.  System modifications that 
suitably reduce their acceleration efficiency are therefore necessary and these are subject to 
the constraints of time variability, just as for our successful invocation of the SDA process at shocks.
This assessment also applies to our subsequent study of Mrk 501, and blazars in general.

\subsection{Application to Mrk 501}
 \label{sec:Mrk501}

\begin{figure}[ht]
\vspace{0.5cm}
\centering
\centerline{\includegraphics[width=8cm]{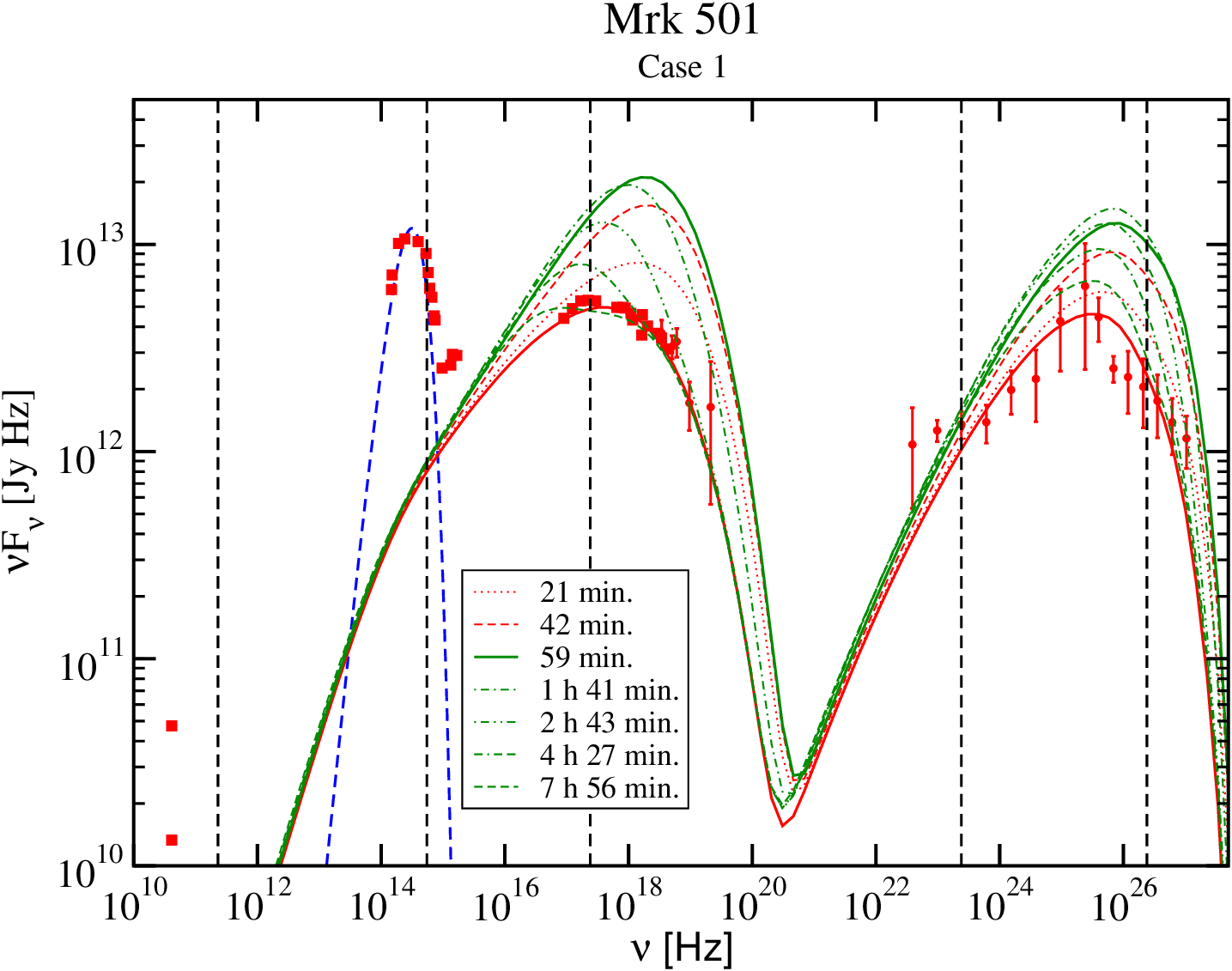}}
\caption{SED of Mrk~501, with data from \cite{Abdo11}. Model curves illustrate the flare simulation for
case 1 (variation of only the electron injection luminosity). Red curves indicate model SEDs during the 
rising, green curves during the decaying part of the flare. The dashed vertical lines indicate the frequencies
at which light curves and hardness-intensity diagrams are extracted (see Table~1).}
\label{Mrk501_SEDs}
\end{figure}

Mrk 501 is one of the archetypal TeV blazars, an HBL at a redshift of $z = 0.034$ \citep[e.g.,][]{Grazian00}, and 
the second extragalactic source detected in VHE $\gamma$-rays by the Whipple Telescope \citep{Quinn96,Bradbury97}. 
The SED of Mrk 501 is typical of HBLs with the synchrotron peak in the X rays, and the SSC (in our leptonic interpretation)
peak located at VHE $\gamma$-rays. The synchrotron component often dominates the total
bolometric output so that the Compton-dominance parameter is $C \lesssim 1$.  The very hard $\gamma$-ray 
spectrum of Mrk 501 was reflected in a non-detection at GeV energies by EGRET, even during VHE $\gamma$-ray flaring episodes 
\citep{Catanese97}, but the improved sensitivity of {\it Fermi}-LAT allowed detailed studies of its GeV spectral 
properties and variability \citep{Abdo11}. Mrk~501 is one of only a handful of blazars from which VHE $\gamma$-ray 
variability on time scales down to a few minutes has been observed \citep{Albert07}. Our study here of the expected 
spectral variability of Mrk 501 in an internal-shock model with consistent particle acceleration generated at shocks
is based on the long-term averaged SED compiled by \cite{Abdo11}, see Fig. \ref{Mrk501_SEDs}.

The SEDs of HBLs are often successfully reproduced by pure SSC models, requiring no external radiation fields
as targets for Compton scattering to produce the $\gamma$-ray emission \citep[e.g.,][]{Ghisellini10}, and the
same holds true for Mrk~501 \citep[e.g.,][]{Petry00}. A quiescent-state SED fit to Mrk~501 with a steady-state
version of our model was already presented in BBS17, and a similar fit with a pure SSC model serves as the 
starting point for our variability study here. Table \ref{Mrk501parameters} lists the parameters used, 
and we remark that they differ somewhat from those chosen in BBS17, most notably by the adoption 
here of a higher magnetic-field strength, but a lower electron injection luminosity.
A noteworthy difference from the case of 3C~279 is the large escape time scale parameter 
$\eta_{\rm esc}$ needed for Mrk 501. This is required by the SED data in order to achieve a cooling break at relatively
low energies, as otherwise the predicted GeV $\gamma$-ray spectrum would be too hard to be consistent
with the {\it Fermi}-LAT spectrum.

\begin{table}[ht]
\caption{\label{Mrk501parameters}Parameters for the quiescent-state model fit to Mrk~501.}
\smallskip
\begin{center}
\begin{tabular}{cc}
\hline
\noalign{\smallskip}
Parameter & Value \\
\noalign{\smallskip}
\hline
\noalign{\smallskip}
Electron injection luminosity & $L_{\rm inj, q} = 1.0 \times 10^{39}$~erg~s$^{-1}$ \\
Emission region size & $\ell_{\rm rad} = 1.5 \times 10^{15}$~cm \\
Jet-frame magnetic field & $B = 0.075$~G \\
Escape time scale parameter & $\eta_{\rm esc} = 1.0 \times 10^3$ \\
PAS m.f.p. low-energy limit & $\eta_1 = 250$ \\
PAS m.f.p. scaling index & $\alpha = 1.5$ \\
Bulk Lorentz factor & $\Gamma^{\ast} \approx \delta = 30$ \\
\end{tabular}
\end{center}
\end{table}

It is well-known that, due to light-travel-time constraints, the minute-scale variability of blazars like Mrk~501
can not be reproduced with a single-emission-zone model. An interpretation of such extreme variability events might
require more complicated geometrical setups, such as the mini-jet-in-jet models of \cite{Giannios10,Nalewajko11}.
The exploration of such scenarios is outside the scope of this paper.  Using our two-zone construction, 
we therefore focus on the more typical flaring
behavior, which is characterized by variability on time scales of $\gtrsim 1$~hour. For a shock speed of $v_{\rm s} = 
0.7c$ and a Doppler factor of $\delta = 30$, this yields a characteristic size of the emission zone of $\ell _{\rm rad}
= 1.5 \times 10^{15}$~cm, which must correspond to the radiative cooling time of the highest energy 
electrons in a viable model to reproduce the $\sim 3-6$~hour variability scales.
The mean free path \teq{\lambda_{\rm pas}(p) = [m_ec^2/eB] \, \eta_1\, (p/m_ec)^{\alpha}} 
for diffusion of particles for our Monte-Carlo shock acceleration simulation yields a good fit
to the Mrk~501 average SED when characterized by $\eta_1 = 250$ and $\alpha = 1.5$. 
With a magnetic field of \teq{B=0.075}G, \teq{m_ec^2/eB = 2.27 \times 10^4}cm. Thus, 
if the characteristic size \teq{\ell_{\rm acc} \equiv \lambda_{\rm pas}(\gammax )} of the 
acceleration zone is set equal to \teq{\ell_{\rm rad}}, the confinement constraint leads to 
a high-energy cut-off in the particle spectrum at $\gammax = 4.1 \times 10^5$. 
In the quiescent-state equilibrium, the magnetization of the emission region is 
$u_B/u_e = 1.8 \times 10^{-3}$.
The solid red curve in Fig. \ref{Mrk501_SEDs} depicts the resulting quiescent-state SED fit. 
Note that the optical flux from Mrk 501 appears to be strongly dominated by the 
contribution of the host galaxy and is unrelated to the jet emission, a common 
presumption for this source; see, e.g., BBS17 and references therein.

Two generic flaring scenarios are addressed in the following exposition, 
with parameters summarized in Table \ref{Mrk501flare_parameters}: (1) A case analogous 
to the 3C~279 Flare C simulation presented in Section \ref{sec:flareC}, changing only the electron injection luminosity
without modifications of the scattering mean free path parameters, and (2) a case similar to the 3C~279 Flare B simulation of
Section \ref{sec:flareB}, where we also change the pitch angle diffusion parameters to produce a harder electron injection
spectrum, in addition to a higher injection luminosity.

\begin{table}[ht]
\caption{\label{Mrk501flare_parameters}Flare-state parameters for variability modeling of Mrk~501.}
\smallskip
\begin{center}
\begin{tabular}{cccc}
\hline
\noalign{\smallskip}
Parameter & Quiescent & Case 1 & Case 2 \\
\noalign{\smallskip}
\hline
\noalign{\smallskip}
$L_{\rm inj}$ [erg~s$^{-1}$] & $1.0 \times 10^{39}$ & $1.0 \times 10^{40}$ & $3.0 \times 10^{39}$  \\
$\eta_1 $ & 250 & 250 & 200 \\
$\alpha$ & 1.5 & 1.5 & 1.4 \\
\end{tabular}
\end{center}
\end{table}

\subsubsection{Case 1: Variation of injection luminosity}
 \label{sec:Mrk501_case1}

In Case 1, we explore a scenario where a strong shock passing through the emission zone enhances only the 
non-thermal electron injection rate by a factor of 10 to $L_{\rm inj, f} = 1.0 \times 10^{40}$~erg~s$^{-1}$, 
but leaves the scattering mean free path parameters unchanged. This is analogous to the Flare C simulation for 3C~279
in Section \ref{sec:flareC}. The resulting snap-shot SEDs are shown in Fig. \ref{Mrk501_SEDs} as red curves during 
the rising phase of the flare and green curves during the decaying portion, respectively. The heavy 
solid green curve indicates the peak of the flare, from which one discerns a slight blue-ward shift 
in the both the synchrotron and SSC peaks; this is caused by progressive acceleration while 
cooling ensues.

\begin{figure}[ht]
\vspace{0.5cm}
\centering
\includegraphics[width=8cm]{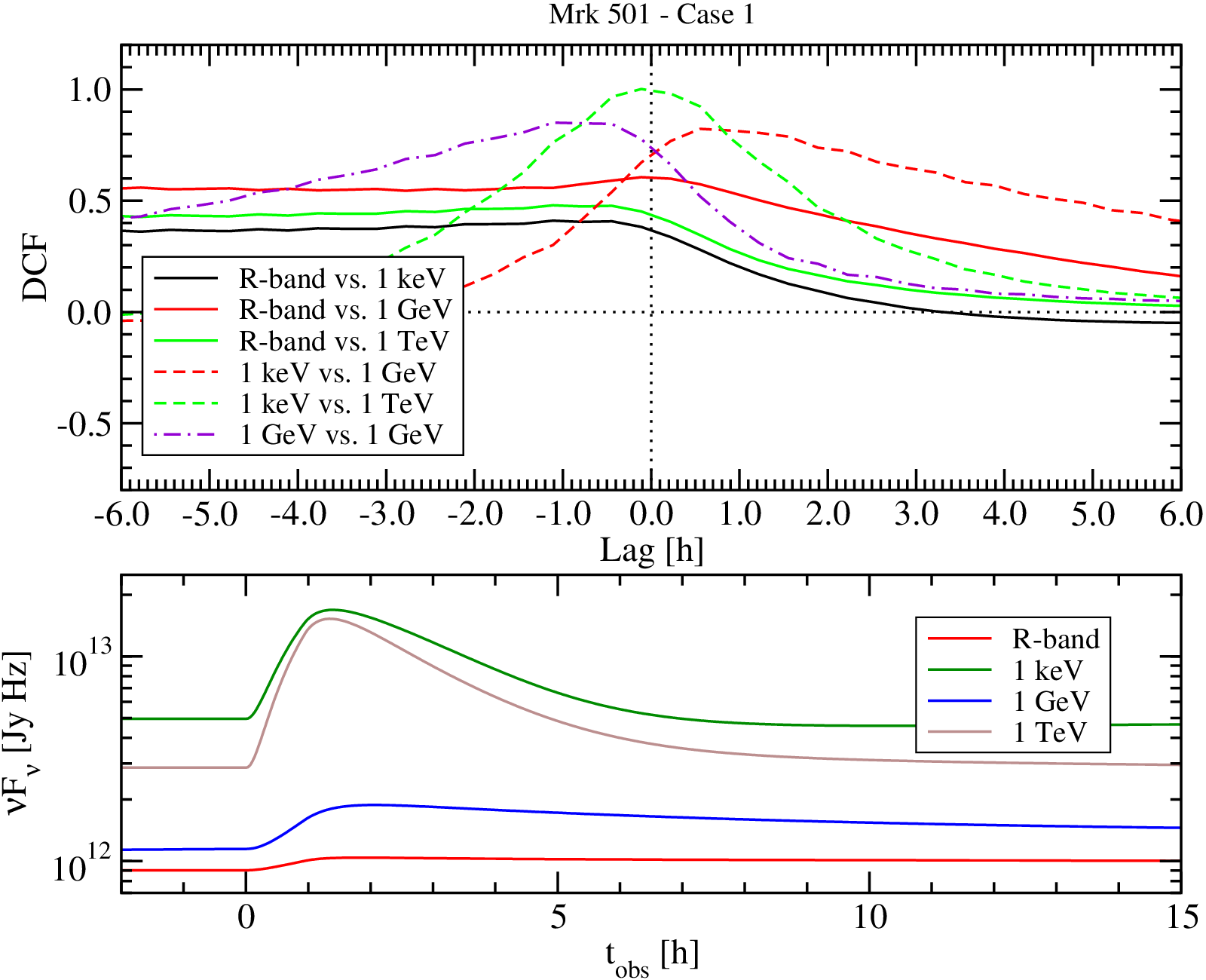}
\caption{{\it Bottom:} Multi-wavelength light curves extracted from the case 1 flare simulation for Mrk~501 
illustrated in Fig. \ref{Mrk501_SEDs}. 
{\it Top:} Discrete cross correlation functions evaluated from the case 1 light curves shown in the bottom panel. }
\label{LC_Mrk501}
\end{figure}   

Fig. \ref{LC_Mrk501} shows the MW light curves for Mrk~501.  Obviously, significant
TeV $\gamma$-ray emission is produced, but the GHz radio band is deep inside the optically-thick part of the 
synchrotron spectrum, with strongly suppressed flux. The radio light curves are therefore ignored in the 
following analysis: it is widely presumed that the radio signal emanates from much larger regions 
of the jet than do the prompt high-energy flares.

Simple scaling arguments based on increases in electron number densities
suggest that one would expect an approximately quadratic 
dependence between the flare amplitudes at the synchrotron and SSC peak frequencies, $\Delta F_{\rm SSC} 
\propto (\Delta F_{\rm syn})^2$, which is obviously not the case in the SEDs shown in Fig. \ref{Mrk501_SEDs}. 
However, as the light curves in the bottom panel of Fig. \ref{LC_Mrk501}
illustrate, we find an approximate scaling 
of $\Delta F_{\rm 1 TeV} \propto (\Delta F_{\rm 1 keV})^{3/2}$ for the TeV vs. 
X-ray light curves (note the logarithmic scaling of the flux axis), 
where the 1 TeV energy is substantially beyond the SSC peak.  Comparing multi-GeV energies,
 an even weaker dependence of the instantaneous flux ratios emerges. This is a result of the
time delay due to the gradual build-up of the synchrotron radiation field and the electron population 
responsible for SSC emission in the GeV band. While the synchrotron flaring amplitude is initially larger than the GeV one, 
the GeV lightcurve decays much more slowly than the synchrotron one 
because it is generated by electrons with lower Lorentz factors.  During the late decay phase, several hours 
after the peak, the GeV $\gamma$-ray flux remains elevated far above the quiescent-state level, while the 
synchrotron flux has essentially decayed back to its quiescent value.

The DCFs in the top panel of Fig. \ref{LC_Mrk501}
illustrate this point further. While the keV and TeV light curves are 
tightly correlated with almost zero time lag, the optical and GeV $\gamma$-ray light curves are delayed 
by $\lesssim 0.5$~hrs with respect to the X-ray and TeV light curves, with the long tail towards negative 
lags resulting from the very slow decay of the optical and GeV light curves. 
The lag and recovery timescales for this case for Mrk 501 are almost an order of magnitude smaller than 
those for our Flare C study for 3C 279.  This is primarily due to the shorter rise time of 
the flares in the case of Mrk 501 as a consequence of the smaller emission region (and, thus,
shorter shock-crossing time), counteracted by the longer radiative cooling time scales for electrons
at the highest energies. The radiative cooling time scale scale is proportional to $\left(B^2 \, [1 + C] 
\gamma_{\rm max}\right)^{-1}$, which is $\sim 2.2$ times larger for Mrk~501 compared to 3C~279 
Flare C.

\begin{figure}[ht]
\vspace{0.5cm}
\centering
\includegraphics[width=8cm]{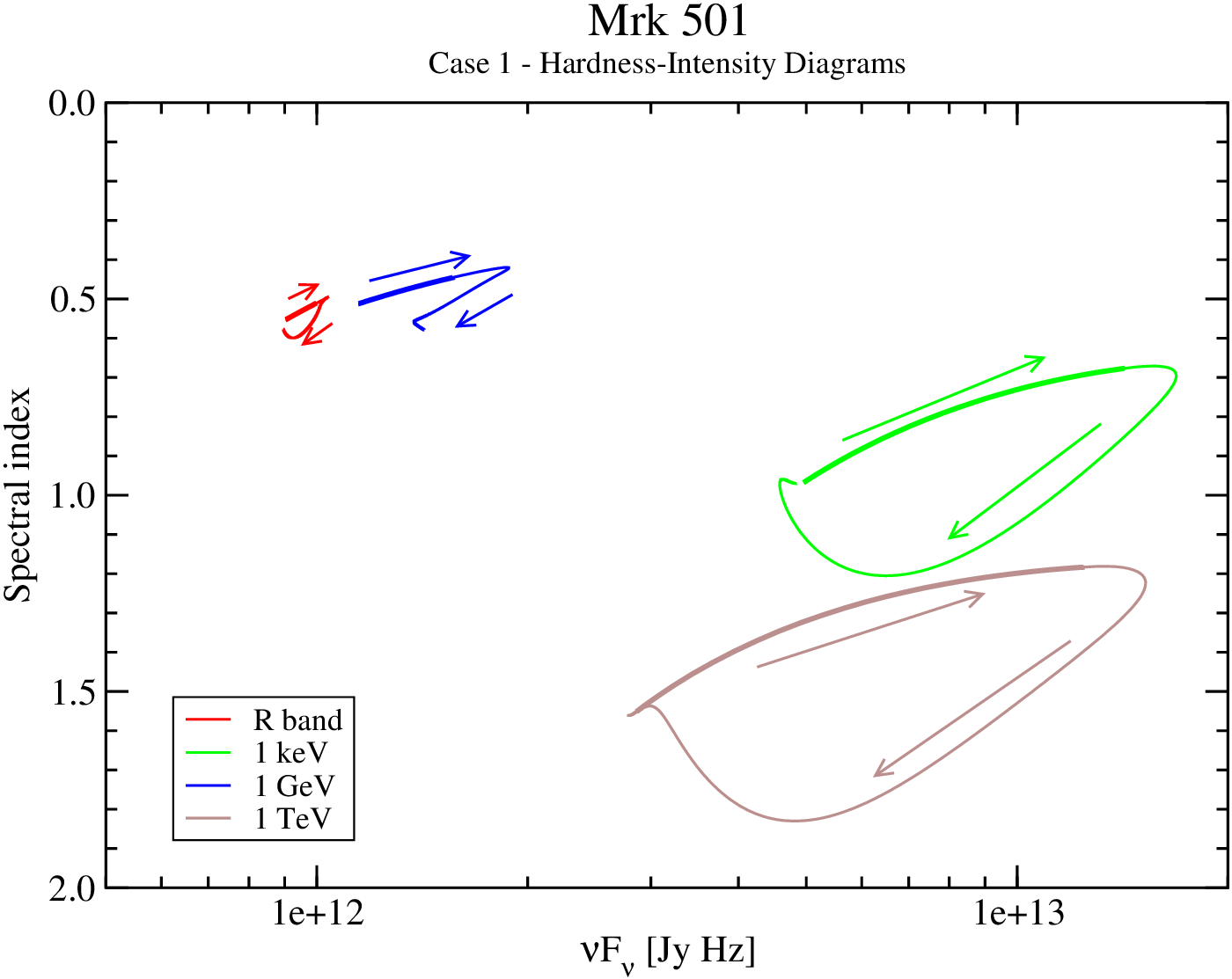}
\caption{Hardness-intensity diagrams for the case 1 flare simulation for Mrk~501 illustrated in Figs. 
\ref{Mrk501_SEDs} --- \ref{LC_Mrk501}. }
\label{HID_Mrk501}
\end{figure}

The hardness-intensity diagrams shown in Fig. \ref{HID_Mrk501} indicate very pronounced clockwise spectral hysteresis
in both the X-ray and TeV $\gamma$-ray bands. The large ``amplitude'' of the hysteresis is a signature of the flux and frequency mobility 
of the synchrotron and SSC peaks around the chosen frequencies.
Such hysteresis has been detected in the case of Mrk~501's
``cousin'', Mrk~421 \citep{Takahashi96}.  In particular, Figs.~6 and~7 of \cite{Tramacere09AA} present HIDs for 
Swift XRT X-ray observations of 5 flares during the April -- July 2006 active period for Mrk 421.
The flares were of sub-hour durations.  Both clockwise and counter-clockwise hysteresis patterns are present at energies \teq{0.2-10}keV 
in the ensemble, and in some cases both directions are realized in a single flare.
Fig.~6 of \cite{Garson10ApJ} displays both clockwise and tilted figure-8 hysteresis in \teq{0.5 - 2}keV Suzaku data 
spanning several hours for modest Mrk 421 activity over 4 days in May 2008, a period when the source did not exhibit strong flares.
More recently, Figs.~12 and 13 of \cite{Abeysekara17ApJ}
display hysteresis profiles for both X-ray and TeV bands on hour-long timescales for two flares in April and May 2014.
The HIDs there exhibit clockwise, counter-clockwise and figure-8-like evolution for the hysteresis in both 
X-ray and VHE $\gamma$-ray bands.  A similar mix is present in the RXTE data for select flares 
of Mrk 421 in Fig.~3 of \cite{Wang18ApJ}.   This presentation of archival RXTE observations of the 
5 brightest blazars includes 3 flares from Mrk 501 that also display a mix of hysteresis directions.
No clear patterns emerge from this dataset, perhaps the most extensive HID information in the literature for Mrk 501. 
At much higher energies, the pronounced TeV-band
hysteresis predicted by our simulation suggests that its detection should be feasible, at least with 
the next-generation IACT facility, the \v{C}erenkov Telescope Array (CTA).

\begin{figure}[ht]
\vspace{0.5cm}
\centering
\centerline{\includegraphics[width=8cm]{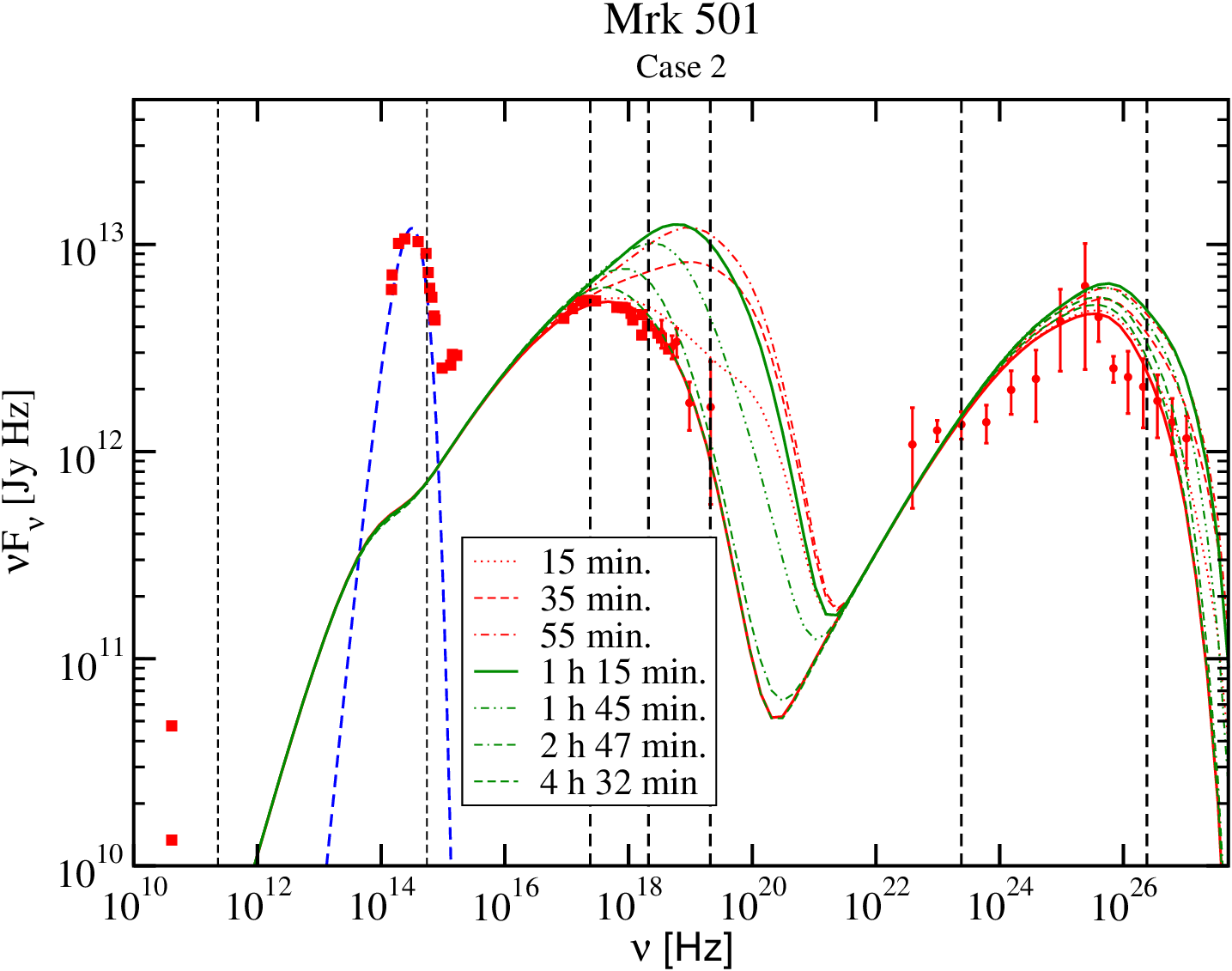}}
\caption{SED of Mrk~501 \citep[data from][]{Abdo11} with model curves for case 2 (changing electron 
luminosity + PAS parameters to produce a harder injection spectrum). Red curves indicate model SEDs 
during the rising, green curves during the decaying part of the flare. The dashed vertical lines 
indicate the frequencies at which light curves and hardness-intensity diagrams are extracted (see Table~1); 
thin lines indicate the radio and optical frequencies which have been ignored in the further analysis.}
\label{Mrk501alpha_SEDs}
\end{figure}

\subsubsection{Case 2: Variation of $L_{\rm inj}$, $\eta_1$, and $\alpha$}
 \label{sec:Mrk501_case2}

The spectral variability of Mrk~501 is peculiar in that it sometimes exhibits extreme shifts of 
the synchrotron (and, to a lesser extend, $\gamma$-ray) peak frequency to higher values during 
flaring states, by more than two orders of magnitude -- see, in particular \cite{Acciari11}. 
Such behavior is obviously not reproduced by our case 1 flare simulations presented above. Therefore, 
as a second test case, we investigated a scenario in which, in addition to an increased electron
injection luminosity, $L_{\rm inj} = 10^{39} \to 3 \times 10^{39}$~erg~s$^{-1}$, the 
pitch angle scattering parameters are also changed, specifically $\eta_1 = 250 \to 200$ 
and $\alpha = 1.5 \to 1.4$.  This yields a smaller diffusive mean free path \teq{\lambda_{\rm pas}}
for electrons at all energies between thermal and the maximum \teq{\gammax m_ec^2},
which has the effect of reducing the acceleration time, rendering 
DSA and SDA more efficient.  Physically, this corresponds to somewhat higher levels of MHD turbulence
on a range of spatial scales. The resulting evolution of the simulated
SEDs is illustrated in Fig. \ref{Mrk501alpha_SEDs}. As for case 1, there is no appreciable radio 
emission from the part of the jet simulated here, and the optical emission is host-galaxy dominated, 
thus showing negligible variability. Therefore, in the following, both the radio and optical light 
curves will not be considered.

\begin{figure}[ht]
\vspace{0.5cm}
\centering
\includegraphics[width=8cm]{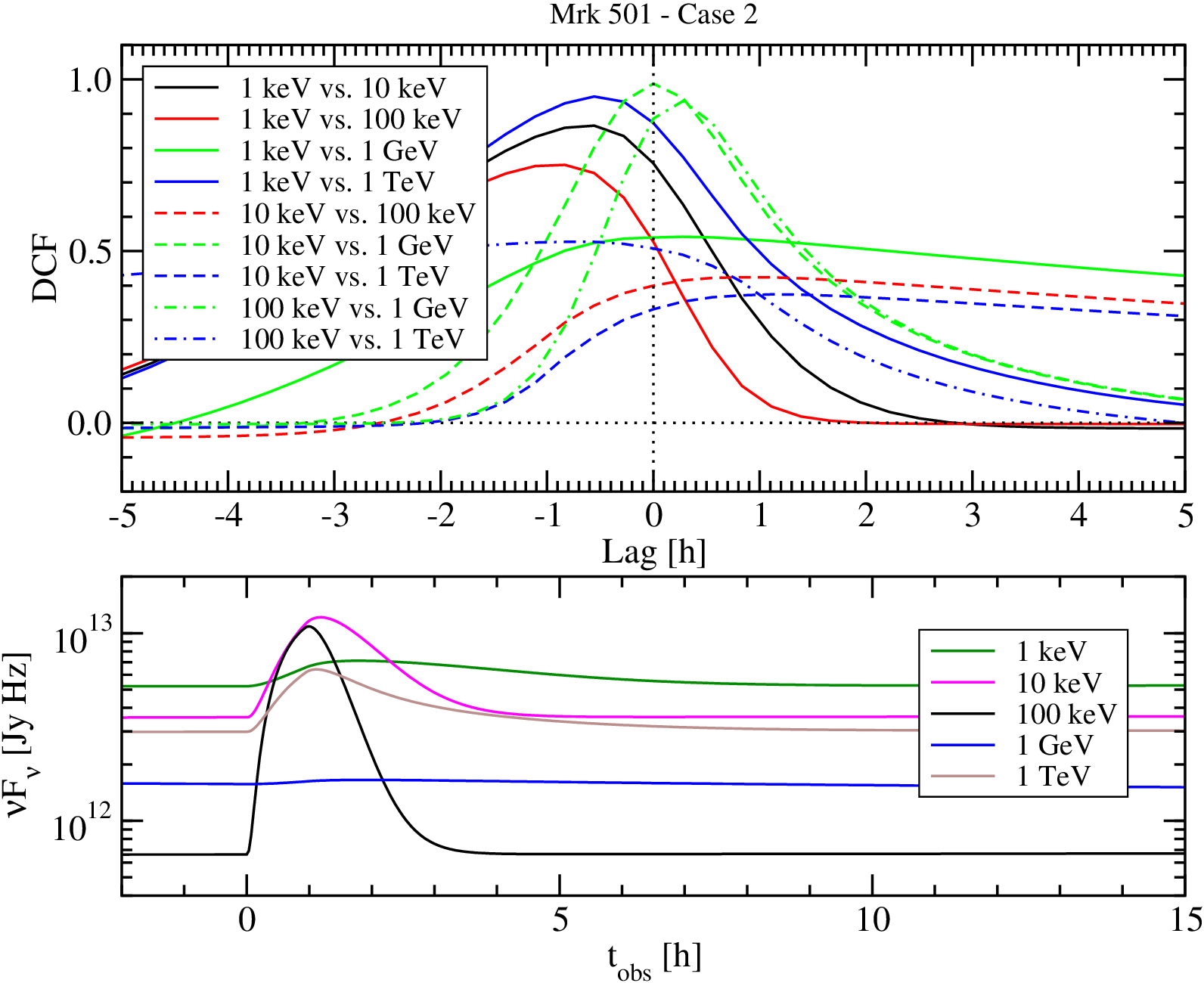}
\caption{{\it Bottom:} Multi-wavelength light curves extracted from the case 2 flare simulation for Mrk~501 
illustrated in Fig. \ref{Mrk501alpha_SEDs}. 
{\it Top:} Discrete cross correlation functions evaluated from the case 2 light curves shown in 
the bottom panel.}
\label{Mrk501alpha_LCs}
\end{figure}

The high-frequency synchrotron spectrum predicted by our case 2 simulation shows the expected extreme 
spectral hardening, with a shift of the synchrotron peak by about a factor of 100, similar to the trend 
observed in \cite{Acciari11}. This hardening is driven by the greater rapidity of the acceleration, 
which moves the maximum Lorentz factor in the model up to \teq{\gammax \approx 1.2 \times 10^6}.
Most of the spectral changes occur beyond the quiescent-state synchrotron 
peak frequency ($\sim 1$~keV), thus resulting in only very moderate variability at 1~keV X-rays. Extreme 
variations, however, occur in the hard X-ray regime at $\sim 10$ -- 100~keV. Therefore, in addition to 
the standard analysis frequencies used for the previous simulations, we extract light curves and 
hardness-intensity diagrams at two additional X-ray frequencies, corresponding to 10~keV and 100~keV. 

The X-ray (1, 10, and 100 keV) and $\gamma$-ray light curves from our case 2 simulation are plotted
in the bottom panel of Fig. \ref{Mrk501alpha_LCs}. 
As expected from the SED evolution (Fig. \ref{Mrk501alpha_SEDs}), the 
variability amplitude is largest at 100 keV and negligible in the {\it Fermi}-LAT regime (1 GeV). 
The light curves also suggest time lags within the X-ray bands, with the high-energy variability 
leading the lower energies. 

The decay time scales of the various X-ray light curves in Fig. \ref{Mrk501alpha_LCs} reflect 
the energy-dependent radiative (synchrotron + SSC) cooling time scales, 
\begin{eqnarray}
   t_{\rm cool}^{\rm obs} & \;\approx\; & \dover{0.9}{1+C} \left( {1 + z \over \delta} \right)^{1/2} \, B_{\rm G}^{-3/2} \,
               E_{\rm keV}^{-1/2} \; {\rm hr} \nonumber\\[-5.5pt]
 \label{eq:tsyn}\\[-5.5pt]
   & \approx & 4 \, E_{\rm keV}^{-1/2} \; {\rm hr} \nonumber
\end{eqnarray}
for the parameters chosen here, where \teq{B_{\rm G}=0.075} is the magnetic field in units of Gauss and 
$C = L_{\rm SSC} / L_{\rm syn} $ ($\sim 1$ for Mrk~501) is the Compton dominance factor, and 
$E_{\rm keV}$ is the synchrotron photon energy under consideration \citep{Takahashi96,Boettcher03}
in units of keV. This yields $t_{\rm cool}^{\rm obs} \approx 4$~hr for 1 keV and $t_{\rm cool}^{\rm obs} 
\approx 0.4$~ hr for 100 keV. The presence of the expected time lags within the X-ray band is confirmed 
by the DCFs plotted in the top panel of Fig. \ref{Mrk501alpha_LCs}. 
However, the lags identified by the DCF are significantly 
shorter than the differences in cooling time scales, because the DCF peak lags are more strongly 
dominated by the relative light curve peak times rather than the decay time scales.

\begin{figure}[ht]
\vspace{0.5cm}
\centering
\includegraphics[width=8cm]{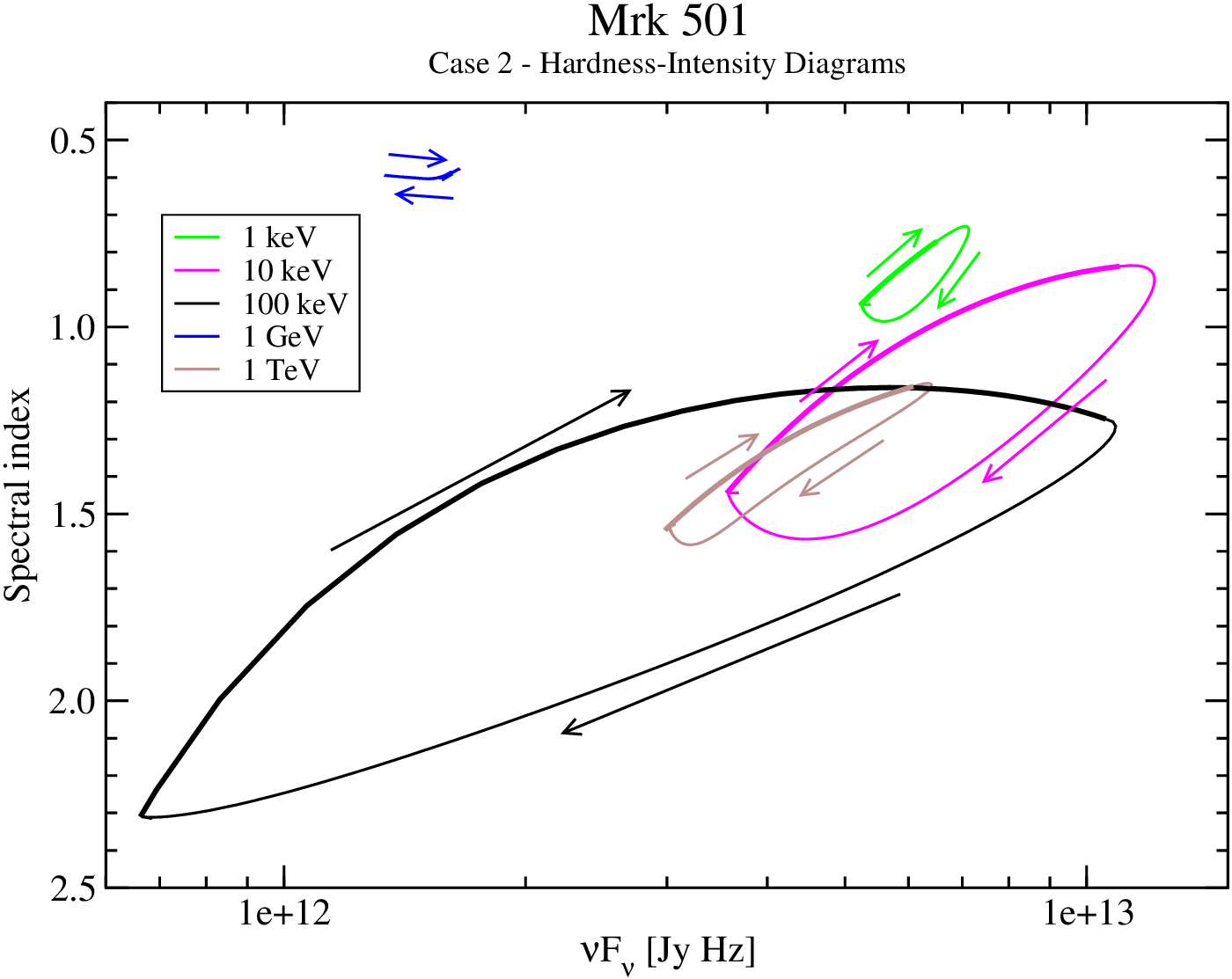}
\caption{Hardness-intensity diagrams extracted from the case 2 flare simulation for Mrk~501 illustrated in Figs. 
\ref{Mrk501alpha_SEDs} --- \ref{Mrk501alpha_LCs}. }
\label{Mrk501alpha_HIDs}
\end{figure}

As illustrated in Fig. \ref{Mrk501alpha_HIDs}, the Case 2 modeling predicts strong spectral hysteresis
in the hard X-ray regime (10~keV and 100~keV), with spectral-index variations of $\Delta\Gamma_{\rm ph}
\sim 0.7$ between the rising and decaying parts of the flare. A sensitive hard X-ray telescope, such 
as NuSTAR, should be able to identify such hysteresis patterns. Extensive NuSTAR observations of Mrk~501
do exist \citep[e.g.,][]{Pandey17,Bhatta18}. However, while a clear harder-when-brighter trend is clearly
seen, e.g., in Fig. 1 of \cite{Bhatta18}, no spectral hysteresis could be clearly identified in the 
\teq{5-70} keV window. In our models, modest spectral hysteresis is also predicted at 1 keV and 1 TeV.

\section{Summary and Discussion}
 \label{sec:discuss}

We present the development of a numerical scheme to couple Monte-Carlo simulations of diffusive shock 
acceleration with time-dependent radiation transfer in an internal-shock, leptonic scenario for blazar flares. 
Our model consists of two zones: A small acceleration zone, in which both DSA and SDA are active, and a larger
radiaton zone, into which shock-accelerated electrons are injected in a time-dependent manner. This
code has been applied to two prototypical blazars: the FSRQ 3C~279 and the HBL Mrk~501. In both
cases, we base our DSA+SDA simulations on a mildly relativistic shock with $v_{\rm s} = 0.7 \, c$ and parameterize 
the pitch-angle scattering mean-free-path of particles as a power law in particle energy, 
$\lambda_{\rm pas} = \eta_1 \, r_g \, (p/m_ec)^{\alpha-1}$, i.e., $\lambda_{\rm pas} \propto 
\gamma^{\alpha}$ at ultra-relativistic energies. As elaborated in our previous work \citep{BBS17}, 
producing synchrotron spectra 
with $\nu F_{\nu}$ peaks in the optical / infrared (as for low-frequency peaked blazars), requires 
a strongly electron energy-dependent $\lambda_{\rm pas}$. Specifically, for 3C~279, we obtain a 
good fit to the average low-state SED using $\eta_1 = 100$ and $\alpha = 3$. For Mrk~501, with 
an average quiescent-state synchrotron peak at $\sim 1$~keV, this requirement is relaxed, and a
good fit can be obtained with $\eta_1 = 250$ and $\alpha = 1.5$.

Multi-wavelength flares of 3C~279 and Mrk~501 have been simulated by changing the particle 
acceleration parameters in a time-dependent way.  We started our explorations 
in Section~\ref{sec:flareC} with Flare C, a somewhat atypical flare of 3C~279 with 
equal flaring amplitude in the synchrotron and external Compton component.
The MW spectrum for this event can be reproduced by simply
invoking a higher rate of particle injection, e.g., due to the shock encountering an over-density 
in the active region of the jet. For this purpose, it has to be postulated that this change in
density does not alter the turbulence characteristics on the shock scale in the sense that $\lambda_{\rm pas}
(p)$ remains unchanged.  The density enhancement would form from MHD substructure on larger, 
super-parsec scales in the jet, and would not be part of shock-associated turbulence.  
In such a scenario, the X-ray and radio variability is expected 
to be delayed with respect to the (simultaneously varying) optical and GeV $\gamma$-ray 
emission by several hours. Moderate spectral hysteresis is predicted for these two bands 
in our modeling, though this is likely undetectable by current 
or near-future instrumentation.

An extreme case of an ``orphan'' $\gamma$-ray flare with strongly increasing $\gamma$-ray flux 
and hardening $\gamma$-ray spectrum, without an accompanying optical flare, requires more dramatic 
changes in the jet.  This provides a more stringent challenge for our two-zone, 
acceleration+radiative dissipation modeling.
In the framework of leptonic blazar models, such events may be plausibly
explained in scenarios where the external target photon field for Compton scattering is temporarily
increased, such as the synchrotron mirror models of \cite{BD98,Tavani15} or the ``Ring of Fire'' model
of \cite{McDonald15,McDonald17}.  In this paper, as an alternative scenario, we explored the question whether such flaring 
behavior can plausibly be reproduced by changing only the particle acceleration characteristics.
The harder $\gamma$-ray spectrum then required a significantly increased particle acceleration 
efficiency, implying a smaller $\lambda_{\rm pas}$ at high energies. This can be achieved with 
smaller values of $\eta_1$ and $\alpha$, indicative of modest increases in turbulence levels in the post-shock
region \citep{SB12}.  

Probes of turbulence changes in blazar shock environs that are independent of this modeling 
are best acquired through optical polarimetry.  Imaging in optical is not possible on the sub-parsec 
scales of the shocks embedded in jets, but time-dependent measures are.  One of the best known 
examples of this for a blazar is actually for 3C 279, detailed in \cite{Abdo2010Nat}, wherein 
a dramatic change in optical polarization signatures accompanied a strong ``orphan'' gamma-ray flare 
measured by {\it Fermi}-LAT, beginning on 18th February, 2009 (MJD 54880).  Data 
collected by the Kanata-TRISPEC and La Palma KVA telescopes indicated a gradual 
change in the polarization angle by \teq{208^{\circ}} that was accompanied by a sharp 
drop in the V-band polarization degree (PD) from around 35\% to around 10\%.  
The polarization degree decline is consistent with an increased level of MHD turbulence
on the light-day scales.  In our interpretation here, it would signal changing conditions 
in the pertinent shocked jet region.  Directly germane to 
the Flare B case study here, Figure~7 of \cite{Hayashida15} exhibits
V-band optical polarization data from the Japanese 
Kanata HOWPol telescope at the level of around 23\% 
just subsequent to Flare B. This would indicate only modest levels of turbulence on 
3--5 light-day lengthscales.

In passing, we remark that while the February 2009 flare 
was almost as strong as our focal 2013 flare, its spectrum in the GeV band was not as flat as 
that of Flare B.  The MW SED exhibited in Figure~2 of \cite{Abdo2010Nat} 
suggests that it would probably be well-modeled with similar parameters to our Flare B example, 
perhaps with slightly lower \teq{\gammax}, and would also require the choice of a low magnetic field.  

An attractive element of our modeling was that the MW spectroscopy and the 
timescales for Flare B could be accommodated with a single set of system parameters.
A notable nuance to this Flare B modeling was that a reduction of the magnetic field
was needed in order to compensate the presence of additional high-energy 
electrons. It is important that after the end of the flaring injection episode, the magnetic field
was required to gradually relax back to restore its higher quiescent-state level to compensate 
for the gradual cooling of the additionally injected high-energy electrons. 
A key question concerns whether such prompt but ephemeral reductions are realistic.  While MHD simulations 
of jets typically only indicate magnetic fluctuations in \teq{\vert \mathbf{B}\vert} by factors of \teq{2-3}, they cannot resolve 
structures on the sub-parsec scales pertinent to this question.  Interestingly, kinetic plasma PIC 
simulations of relativistic shocks do see strong magnetic field strength contrasts, by factors of \teq{10-20} in 
Weibel-instability-generated
turbulence \citep[e.g.,][]{Sironi13ApJ}. However, these variations are on the extremely small inertial 
scales \teq{c/\omega_p \sim 10^3-10^6}~cm that are not germane to the high energy electrons that emit 
optical synchrotron and inverse Compton gamma-rays.  

Yet it is interesting to note that magnetometer measurements
of active regions just downstream of non-relativistic interplanetary shocks in the solar wind 
do exhibit sharp magnetic rarefactions, albeit by factors of around \teq{1.5-2} in field strength
\citep[e.g., see Ulysses B-field data streams in][]{BOEF97}, followed by some recovery.  The magnetic field in these 
heliospheric rarefactions also changes direction, perhaps in a manner broadly consistent 
with what is invoked to explain the 3C 279 polarization angle swing highlighted in \cite{Abdo2010Nat}.
On the basis of these varied pieces of information, it appears that our choice of a reduction 
in \teq{\vert \mathbf{B} \vert} by around a factor of 9 is not particularly concerning, 
though it is difficult to draw tight conclusions pertaining to its appropriateness
without further observational information.  Results from a future gamma-ray polarimeter, such as 
the planned AMEGO\footnote{see {\tt https://asd.gsfc.nasa.gov/amego/index.html}.}
and e-ASTROGAM \citep[see][]{DeAngelis17} missions, could well enlighten the picture.

For Mrk~501 we also started with a simple scenario of increasing the number of injected 
electrons without changing the turbulence characteristics, i.e., leaving $\lambda_{\rm pas}
(\gamma)$ unchanged. Such a scenario leads to correlated variability across the electromagnetic
spectrum (except for the optical, which is host-galaxy dominated), with the GeV $\gamma$-ray 
flux variations lagging behind the simultaneously-varying keV X-ray and TeV fluxes by $\sim 
1$~hour. Significant spectral hysteresis in the X-ray and TeV bands was predicted. 

Mrk~501 sometimes exhibits a significant shift of the synchrotron peak frequency to higher 
values during flaring states \citep[e.g.,][]{Acciari11}. In order to reproduce such a scenario,
we ran a second simulation, invoking an increased particle acceleration efficiency through 
a decreasing $\lambda_{\rm pas}$ at high energies (changing $\eta_1 = 250 \to 200$ and 
$\alpha = 1.5 \to 1.4$). Such changes could be indicative of modest increases in turbulence levels 
in the post-shock region. This case study approximately reproduces the significant synchrotron peak 
shift observed by \cite{Acciari11}, and predicts strong hard X-ray spectral hysteresis,
potentially detectable by NuSTAR, and time lags between different X-ray energy ranges
(1 -- 100~keV) of the order of $\lesssim 1$~hour. Importantly, the time lags 
identified by the discrete correlation function are significantly shorter than the 
differences in radiative cooling time scales, which govern primarily the decay time
scales of flares at different energies. Several authors \citep[e.g.,][]{Takahashi96,Boettcher03} 
have suggested that measured inter-band time lags can be used to estimate the strength of the magnetic field
under the assumption that these time lags reflect differences in the radiative cooling time
scales. Our study illustrates that this protocol is at risk of under-estimating the actual radiative cooling time
scales and, thus, over-estimate the magnetic field by a factor of a few. Careful investigation 
via temporal correlation functions can ameliorate this complication.

This fairly diverse selection of flare case studies clearly highlights the broad viability of 
our approach of combining particle acceleration and MW emission simulations 
in a two-zone construction to modeling quiescent and flaring phases of the 
blazars 3C 279 and Mrk 501.  Moreover, it illustrates the richness in time dependent 
information delivered by such an integrated theory analysis, motivating more intensive observational 
scrutiny by multi-wavelength campaigns during active phases of bright blazars.  We anticipate that 
during the CTA era, it will prove possible to perform incisive 
diagnostics into the competition between acceleration and radiative cooling that have a
broader scope than just probing simple temporal injection profiles \teq{Q_e(\gamma_e, \, t)}
like those employed in this investigation. 

\acknowledgements{We thank the anonymous referee and Anita Reimer 
for suggestions helping to improve the presentation.
The work of M. B\"ottcher is supported by the South African 
Research Chairs Initiative (grant no. 64789) of the Department of Science and 
Innovation and the National Research Foundation\footnote{Any opinion, finding 
and conclusion or recommendation expressed in this material is that of the authors 
and the NRF does not accept any liability in this regard.} of South Africa.
MGB and MB are grateful to NASA for partial support for early parts of this research 
program through the Astrophysics Theory Program, grant NNX10AC79G.
MGB is also grateful for support from the NASA {\it Fermi} 
Guest Investigator Program through grant 80NSSC18K1711.}

\end{document}